\documentclass[useAMS,usenatbib]{mnras}
\usepackage{lscape,graphicx,amssymb,aas_macros,bm}
\usepackage{amsmath} 
\usepackage{indentfirst} 
\usepackage{pdflscape} 
\usepackage{placeins} 
\usepackage{nicefrac} 
\usepackage{afterpage} 
\usepackage{booktabs} 
\usepackage{morefloats} 
\usepackage{color} 
\usepackage{float,subfloat}
\usepackage[bottom]{footmisc}

\newcommand{\apl}{\:^{<}_{\sim}\:}
\newcommand{\cmjj}{\mbox{${\rm cm^{-2}}$}}
\newcommand{\cmjjj}{\mbox{${\rm cm^{-3}}$}}
\newcommand{\etal}{et al.}
\newcommand{\HI}{{\mbox{H\,{\scriptsize I}}}}

\newcommand{\kms}{\mbox{km\ s${^{-1}}$}}

\newcommand{\FeII}{{\mbox{Fe\,{\scriptsize II}}}}
\newcommand{\MgII}{{\mbox{Mg\,{\scriptsize II}}}}

\newcommand{\SiII}{{\mbox{Si\,{\scriptsize II}}}}
\newcommand{\CIII}{{\mbox{C\,{\scriptsize III}}}}
\newcommand{\CII}{{\mbox{C\,{\scriptsize II}}}}
\newcommand{\SiIII}{{\mbox{Si\,{\scriptsize III}}}}
\newcommand{\OVI}{{\mbox{O\,{\scriptsize VI}}}}

\newcommand{\msun}{{\mbox{M$_{\odot}$}}}

\title[The Cosmic Ultraviolet Baryon Survey (CUBS) III]{The Cosmic Ultraviolet Baryon Survey (CUBS) - III. Physical properties and elemental abundances of Lyman limit systems at $z<1$}

\author[Zahedy and the CUBS team]{Fakhri S. Zahedy$^{1}$\thanks{E-mail: fzahedy@carnegiescience.edu}, 
Hsiao-Wen Chen$^{2}$, Thomas M.\ Cooper$^{1}$, Erin Boettcher$^{2}$, Sean D.\ Johnson$^{3}$, \newauthor Gwen C.\ Rudie$^{1}$, Mandy C.\ Chen$^{2}$, Sebastiano Cantalupo$^{4,5}$, 
Kathy L.\ Cooksey$^{6}$, Claude-Andr\'e  \newauthor Faucher-Gigu\`ere$^{7}$, Jenny E.\ Greene$^{8}$, Sebastian Lopez$^{9}$,  John S.\ Mulchaey$^{1}$, Steven V.\ Penton$^{10}$, \newauthor Patrick Petitjean$^{11}$, Mary E.\ Putman$^{12}$, Marc Rafelski$^{13,14}$, Michael Rauch$^{1}$,  Joop Schaye$^{15}$, \newauthor Robert A.\ Simcoe$^{16}$, and Gregory L.\ Walth$^{1}$\\
$^{1}$The Observatories of the Carnegie Institution for Science, 813 Santa Barbara Street, Pasadena, CA 91101, USA \\
$^{2}$Department of Astronomy \& Astrophysics, The University of Chicago, Chicago, IL 60637, USA \\
$^{3}$Department of Astronomy, University of Michigan, Ann Arbor, MI 48109, USA \\
$^{4}$Department of Physics, ETH Wolfgang$-$Pauli$-$Strasse 27, 8093, CH-8093 Z\"urich, Switzerland \\
$^{5}$Department of Physics, University of Milan Bicocca, Piazza della Scienza 3, 20126 Milano, Italy\\
$^{6}$Department of Physics and Astronomy, University of Hawaii at Hilo, Hilo, HI 96720, USA\\
$^{7}$Department of Physics \& Astronomy and Center for Interdisciplinary Exploration and Research in Astrophysics (CIERA),\\ Northwestern University, 1800 Sherman Ave, Evanston, IL 60201, USA \\ Department of Physics and Astronomy, Northwestern University, 2145 Sheridan Road, Evanston, IL 60647, USA \\
$^{8}$Department of Astrophysical Sciences, Princeton University, Princeton, NJ 08544, USA \\
$^{9}$Departamento de Astronom\'ia, Universidad de Chile, Casilla 36-D, Santiago, Chile\\
$^{10}$Laboratory For Atmospheric and Space Physics, University of Colorado, Boulder, CO 80303, USA\\
$^{11}$Institut d’Astrophysique de Paris, CNRS-SU, UMR 7095, 98bis bd Arago, Paris F-75014, France \\
$^{12}$Department of Astronomy, Columbia University, New York, NY 10027, USA\\
$^{13}$Space Telescope Science Institute, Baltimore, MD 21218, USA \\
$^{14}$Department of Physics \& Astronomy, Johns Hopkins University, Baltimore, MD 21218, USA\\
$^{15}$Leiden Observatory, Leiden University, PO Box 9513, NL-2300 RA Leiden, the Netherlands\\
$^{16}$MIT-Kavli Institute for Astrophysics and Space Research; 77 Massachusetts Ave., Cambridge, MA 02139, USA \\
}

\begin{document}

\pagerange{\pageref{firstpage}--\pageref{lastpage}} \pubyear{2021}

\maketitle

\label{firstpage}

\begin{abstract}
 
We present a systematic investigation of physical conditions and elemental abundances in four optically thick 
Lyman-limit systems (LLSs) at $z=0.36-0.6$ discovered within the Cosmic Ultraviolet Baryon Survey
(CUBS). Because intervening LLSs at $z<1$ suppress FUV light from background QSOs, an unbiased search of
these absorbers requires a NUV-selected QSO sample, as achieved by CUBS.
CUBS LLSs exhibit multi-component kinematic structure and a complex mix of multiphase gas, with
associated metal transitions from multiple ionization states such as
\CII, \CIII, \ion{N}{III}, \MgII, \SiII, \SiIII, \ion{O}{II}, \ion{O}{III}, \OVI, and \ion{Fe}{II} absorption that span several hundred \kms\ in line-of-sight velocity. 
Specifically, higher column density components (log\,$N$(HI)/\cmjj$ \gtrsim16$) in all four absorbers comprise dynamically cool gas with $\langle T \rangle =(2\pm1) \times10^4\,$K and modest non-thermal broadening of  $\langle b_\mathrm{nt} \rangle =5\pm3\,$\kms. 
The high quality of the QSO absorption spectra allows us to infer the physical conditions of the gas, using a detailed ionization modeling that takes into account the resolved component structures of \ion{H}{I} and metal transitions. The range of inferred gas densities indicates that these absorbers consist of spatially compact clouds with a median line-of-sight thickness of $160^{+140}_{-50}$ pc. While obtaining robust metallicity constraints for the low-density, highly ionized phase remains challenging due to the uncertain $N\mathrm{(\ion{H}{I})}$, we
demonstrate that the cool-phase gas in LLSs has a median metallicity of  $\mathrm{[\alpha/H]_{1/2}}=-0.7^{+0.1}_{-0.2}$, with a 16-84 percentile range of $\mathrm{[\alpha/H]}=(-1.3,-0.1)$. Furthermore, the wide range of inferred elemental abundance ratios ($\mathrm{[C/\alpha]}$, $\mathrm{[N/\alpha]}$, and $\mathrm{[Fe/\alpha]}$) indicate a diversity of chemical enrichment histories. Combining the absorption data with deep galaxy survey data characterizing the galaxy environment of these absorbers, we discuss the physical connection between star-forming regions in galaxies and diffuse gas associated with optically thick absorption systems in the $z<1$ circumgalactic medium. 
\end{abstract}


\section{Introduction}
\label{section:introduction}
 
The extended gaseous haloes surrounding galaxies, commonly known as the circumgalactic medium (CGM),
retain a sensitive record of different baryonic processes that drive galaxy evolution 
(see recent reviews by Tumlinson \etal\ 2017; Chen 2017). In the standard picture of galaxy evolution, galaxies grow inside dark-matter 
haloes by accreting gas from the intergalactic medium (IGM), which cools and condenses as it flows toward galaxies,
eventually forming stars. Star formation and other physical processes in galaxies can in turn drive energetic 
feedback that regulates future accretion. 

For over three decades, QSO absorption spectroscopy has provided a
powerful tool to detect and characterize diffuse gas in both the CGM and IGM, which is challenging to detect 
in emission beyond the local Universe (e.g., Fern{\'a}ndez \etal\ 2016).  
Cosmological hydrodynamical simulations demonstrate that strong \ion{H}{I} absorption systems with log\,$N$(HI)/\cmjj$ \gtrsim16$
are strongly correlated with the presence of nearby galaxy haloes at both low and high redshifts
(e.g., Faucher-Gigu{\`e}re \& Kere{\v{s}} 2011; van de Voort \etal\ 2012; Rahmati \& Schaye 2014; Hafen \etal\ 2017). Specifically, optically thick Lyman limit systems (LLSs;  log\,$N$(HI)/\cmjj$ \gtrsim17.2$)
trace overdense structures (e.g., Cooper \etal\ 2015; Faucher-Gigu{\`e}re \etal\ 2015; Fumagalli \etal\ 2016) within galaxy haloes and are potentially useful tracers of both cosmological inflows and galactic outflows. 

An often-used diagnostic of the origin of CGM/IGM gas is the gas metallicity (e.g., Pei \& Fall 1995; Pettini 2001; Schaye \etal\ 2003; Simcoe \etal\ 2004; Lehner \etal\ 2013; Cooke \etal\ 2017; Zahedy \etal\ 2019a; Sameer \etal\ 2021), which quantifies the overall 
heavy element enrichment level of the gas. Galaxies in the local Universe are known to follow a mass-metallicity relation: higher mass galaxies tend to also exhibit both higher stellar and gas-phase metallicities (e.g., Tremonti \etal\ 2004; Gallazzi \etal\ 2005).  
It is natural to expect gas ejected from more massive galaxies to have higher metallicities than gas stripped from low-mass satellite galaxies.
On the other hand, freshly accreted gas from the IGM should have an even lower metallicity, reflecting its chemically primitive nature (e.g., van de Voort \& Schaye 2012; Ford \etal\ 2014; Angl{\'e}s-Alc{\'a}zar \etal\ 2017; Hafen \etal\ 2019; Zahedy \etal\ 2019b). 

However, these simple expectations are complicated by our lack of understanding of the physical mechanisms responsible for transporting heavy elements out of/onto galaxies, a fact that is further highlighted by recent observational studies. Investigating the gaseous haloes of $z\approx0.5$ massive quiescent galaxies with log\,$M_\mathrm{stellar}/\mathrm{M}_\odot>11$,  Zahedy \etal\ (2019a) found that both gas density and metallicity in the cool ($T\sim10^4$ K) CGM can vary by more than an order of magnitude within individual haloes. These findings underscore that the CGM comprises a mixture of gases with diverse physical origins. Furthermore, the large fluctuations of gas metallicity observed within individual haloes are in line with expectations for a medium in which chemical mixing is an inefficient process (e.g., Scalo \& Elmegreen 2004; Schaye \etal\ 2007). For that reason, gas metallicity alone is an incomplete diagnostic of the origins of chemically enriched gas in the gaseous haloes (e.g., Hafen \etal\ 2019). 

In contrast, the elemental abundance pattern of the gas provides an archaeological record of different sources of heavy element production. For instance, $\alpha$-process elements (such as Mg, O, Si, and S) are predominantly synthesized in massive stars and their ensuing core-collapse supernovae (SNe CC). In contrast, Fe-peak elements like Fe, Ni, and Mn are produced in large quantities by Type Ia SNe, which occur over longer timescales in galaxies (e.g., Tsujimoto \etal\ 1995; Nomoto \etal\ 2006; Kobayashi \etal\ 2020a). Therefore, the observed abundance ratio of $\mathrm{[Fe/\alpha]}$ constrains the relative contributions of SNe CC and SNe Ia into the chemical enrichment history and offers a powerful constraint for the origin of chemically enriched gas around galaxies (e.g., Lu \etal\ 1996; Pettini \etal\ 2002; Zahedy \etal\ 2016; 2017; Boettcher \etal\ 2020). A joint absorber and galaxy study that combines observational constraints on absolute gas metallicity and relative elemental abundance ratios is necessary to unveil the physical connection between stellar populations and star-forming regions in galaxies and diffuse gas in the CGM/IGM. 

To better understand the co-evolution of galaxies and diffuse CGM/IGM gas, we are
conducting the Cosmic Ultraviolet Baryon Survey (CUBS). CUBS is based on a
large {\it Hubble Space Telescope} ({\it HST}) Cycle 25 General
Observer Program (GO-CUBS; ${\rm PID}=15163$; PI: Chen) to obtain
high-quality FUV spectra of 15 UV bright QSOs. 
Our survey aims to map diffuse gas at $z\apl 1$ using
QSO absorption-line spectroscopy, which is complemented 
by deep galaxy survey data (Chen \etal\ 2020). 
CUBS will establish a large sample of galaxies and absorbers at $z<1$,
and enable systematic studies of the  co-evolution of galaxies and their CGM at an epoch of precipitous decline in the cosmic star-formation rate density (SFRD), offering critical insights into how the complex interplay of gas accretion and outflows govern the mass assembly of galaxies over cosmic time. 

We presented the initial results from the CUBS program and reported the discovery of five
new LLSs with log\,$N$(HI)/\cmjj$ \gtrsim17.2$ within our survey footprint in Chen \etal\ (2020; hereafter Paper I). One of the five systems is an $\mathrm{H_2}$-bearing damped Ly$\alpha$ (DLA) system situated in the vicinity of a massive elliptical galaxy (Boettcher \etal\ 2020; hereafter Paper II).  The wide range of galaxy properties observed around these LLSs
demonstrates that optically thick LLSs are associated with a wide variety of galaxy environments, and the observed variations in ionic column density ratios suggest that they likely trace a broad range of gas metallicities. In this paper, we expand our investigation with detailed absorption-line and ionization analyses of both \ion{H}{I} and metal absorbers in these systems, in order to characterize the ionization states, metallicities, and elemental abundance ratios in these $z<1$ LLSs. We also incorporate the known galaxy environments of these absorbers and discuss the possible physical connection between stellar populations and star-forming regions in galaxies and diffuse gas associated with these LLSs. 
 
 The paper is organized as follows. We summarize the CUBS LLS dataset and relevant data reduction steps in Section 2. 
 In Section 3, we describe the absorption-line analysis and ionization modeling. We then 
discuss the overall physical properties and metal enrichment of the LLSs in Section 4. 
 Finally, we discuss the implications of the results in Section 5, and summarize our findings and present the conclusions in Section 6. 
 In addition, a discussion of the analysis results for individual systems is presented separately in Appendix A. 
 A standard $\Lambda$ cosmology is adopted throughout the paper, with $\Omega_M$ = 0.3,
  $\Omega_\Lambda$=0.7, and a Hubble constant of $H_{\rm 0} = 70\rm \,km.\,s^{-1}\,Mpc^{-1}$.
 We adopt the solar chemical abundance pattern from Asplund \etal\ (2009), and all distances and sizes are 
 physical/proper distances.

\section{Data}

Here we summarize the spectroscopic observations and data reduction of background QSOs relevant to this study. 
A full discussion of the CUBS program design and sample selection, and an overview of the accompanying deep galaxy survey were presented in Paper I. Designed to allow an unbiased study of the $z<1$ CGM/IGM, CUBS consists of 15 QSOs at $z_\mathrm{QSO}>0.8$ which are bright in the near-UV (NUV; $1770-2730$ \AA) channel of {\it GALEX}, $\mathrm{NUV}_{AB}<17$. The NUV magnitude-limited selection ensures that the CUBS sample is not biased against detecting partial/full LLSs at $z<0.9$, which would introduce significant flux attenuation in the {\it GALEX} far-UV (FUV; $1350-1780$ \AA) bandpass. 

Medium-resolution FUV spectra of the 15 QSOs were obtained using the {\it HST}'s  Cosmic Origins Spectrograph (COS; Green \etal\ 2012) between 2018 and 2019, as part of the {\it HST} Cycle 25 (PID=15163; PI: Chen). The COS observations use the G130M and G160M gratings and multiple central wavelength settings, which offer a contiguous spectral coverage from $\lambda\approx1100$ \AA\  to $\lambda \approx 1800$ \AA\ at a spectral resolution of ${\rm FWHM}\approx18-20\, \kms$. As described in Paper I, the pipeline-reduced COS data were first downloaded from the {\it HST} archive, and then processed further using a custom software. The final UV spectra of the CUBS QSOs have a median signal-to-noise ratio of S/N $\approx 12-31$ per resolution element over their full spectral range.

We complement the FUV spectra of the CUBS QSOs with high-resolution optical Echelle spectra obtained using the MIKE spectrograph on the Magellan II (Clay) Telescope (Bernstein \etal\ 2003), which provide additional spectral coverage in the optical ($\lambda=3300-9300$ \AA) at a spectral resolution of between ${\rm FWHM}\approx7\, \kms$ and ${\rm FWHM}\approx10\, \kms$. The MIKE observations were conducted as filler targets within a number of different observing programs carried out between 2017 September and 2019 October. The readers are referred to Paper I for a discussion on the data reduction of the MIKE data. By offering a factor of two increase in spectral resolution compared to our COS FUV spectra, the MIKE data play an important role in guiding subsequent absorption analyses.  

As described in Paper I, the search for LLSs in CUBS yielded five new optically thick absorbers with mean opacities at the Lyman limit of $\tau_{912}\gtrsim1$ (equivalent to  log\,$N$(HI)/\cmjj$ \gtrsim17.2$)  at  $z_\mathrm{abs}<1$. This search was based on the apparent flux decrement at the Lyman limit and corroborated by the presence of other associated Lyman series transitions. The optically thick absorber sample includes one damped Ly$\alpha$ (DLA) system with log\,$N$(HI)/\cmjj$ >20$ at $z_\mathrm{abs}=0.57$, which was analyzed in detail in Paper II. In this work, we present a detailed analysis of the remaining four LLSs. 

\section{Analysis}

A primary goal of the current work is to investigate the thermal state, ionization, and elemental abundances in an unbiased sample of $z<1$ LLSs. As described in Paper I, all four LLSs exhibit multicomponent kinematic structures and associated absorption from metal ions probing a wide range of ionization states. Constraining the physical properties of these optically thick absorbers requires 
(1) knowledge of the ionic column densities and line widths of various observed metal absorption features, and
(2) a careful ionization modeling that considers a wide range of possible physical conditions (e.g., density, metallicity, and abundance ratios) of the gas. Here we describe the two stages of the analysis. 
 
\subsection{Absorption Analysis}

As shown in Paper I, the COS FUV and MIKE optical spectra of the QSOs resolve each LLS into multiple distinct kinematic components of varying absorption strengths. The resolved component structures of both \ion{H}{I} and associated metals enable accurate measurements of both absorption linewidths and the ionic column densities. The $N\mathrm{(\ion{H}{I})}$ measurements for the LLS sample were discussed in detail in Paper I, and here we focus on the absorption analysis of the various associated metal species. We perform a Voigt profile analysis to constrain the properties of individual components in each absorber, using a custom software described in Zahedy \etal\ (2019a). For each ionic species, a kinematic component is fully defined by three parameters: the velocity shift $dv_c$ relative to the line centroid of the strongest \ion{H}{I} component of the LLS, the column density log\,$N_c$, and the Doppler parameter $b_c$. We first generate a theoretical spectrum using the fewest kinematic components required to fit the absorption profile well. Next, this theoretical spectrum is convolved by the known line spread function (LSF) of the relevant spectrograph and binned to match the pixel resolution of the data. Finally, we compare the resulting simulated profile to the data and determine the best-fit model using a $\chi^2$minimization routine. 

To estimate uncertainties in the model, we construct a marginalized posterior probability distribution function (PDF) for each model parameter based on a Markov Chain Monte Carlo (MCMC) analysis done using the \textsc{emcee} package (Foreman-Mackey \etal\ 2013). The MCMC analysis enables us to account for correlated errors in blended and/or saturated absorption components, which would result in underestimated parameter uncertainties if unaccounted for. Each MCMC run uses an ensemble of 250 walkers, with 500 steps performed by each individual walker. To speed up convergence, the walkers are seeded in a tiny region of the parameter space surrounding the minimum $\chi^2$ solution. 

For each LLS, we perform the Voigt profile analysis on all observed transitions from the following list, ordered by their increasing rest wavelength:
\ion{O}{IV} $\lambda787$, \ion{O}{II} $\lambda832$, \ion{O}{III} $\lambda832$, \ion{O}{II} $\lambda833$, \ion{O}{II} $\lambda834$, \ion{S}{II} $\lambda906$, 
\ion{O}{I} $\lambda971$, \ion{C}{III} $\lambda977$, \ion{O}{I} $\lambda988$, \ion{N}{III} $\lambda989$, \ion{S}{III} $\lambda1012$, the \ion{O}{VI} $\lambda\lambda1031,1037$ doublet, \ion{C}{II} $\lambda1036$, \ion{N}{II} $\lambda1083$, \ion{Fe}{III} $\lambda1122$, \ion{Fe}{II} $\lambda1144$, \ion{Si}{II} $\lambda1190$, \ion{Si}{II} $\lambda1193$, \ion{Si}{III} $\lambda1206$, the \ion{N}{V} $\lambda\lambda1238,1242$ doublet, \ion{Si}{II} $\lambda1260$, \ion{O}{I} $\lambda1302$, \ion{Fe}{II}\,$\lambda2344$, \ion{Fe}{II}\,$\lambda2374$, \ion{Fe}{II}\,$\lambda2382$, \ion{Fe}{II}\,$\lambda2586$, \ion{Fe}{II}\,$\lambda2600$, the \ion{Mg}{II}\,$\lambda\lambda2796, 2803$ doublet, and \ion{Mg}{I}\,$\lambda2852$. To reduce the number of free parameters, we require different transitions of the same ion (e.g., \ion{Fe}{II} $\lambda2382$ and \ion{Fe}{II} $\lambda2600$) to share identical Voigt profile parameters. We also impose a common kinematic structure (number of components and their centroids) among \ion{H}{I}, low ions (e.g., \ion{Fe}{II}, \ion{Mg}{II}), and intermediate-ionization species (e.g., \ion{C}{III}, \ion{Si}{III}). This choice is motivated by the excellent agreement among the absorption profiles of various low- and intermediate-ionization species (see Paper I for a full discussion). An exception from this requirement is applied to the high-ionization \ion{O}{VI} $\lambda\lambda$1031,1037 doublet, which frequently exhibits a distinct kinematic structure from lower-ionization species (e.g., Savage \etal\ 2010; Zahedy \etal\ 2019a; 2020; Rudie \etal\ 2019). The different kinematic structure of  \ion{O}{VI} suggests that it arises in a distinct and more highly ionized gas phase. We analyze the \ion{O}{VI} absorption profiles independently and separately from the lower-ionization species, without imposing a common kinematic structure between them. In most cases, the \ion{H}{I} column density associated with a more highly ionized gas phase detected in \ion{O}{VI} is significantly ($\gtrsim 1-2$ dex) lower than the $N\mathrm{(\ion{H}{I})}$ of the low-ionization phase (e.g., Sameer \etal\ 2021; Haislmaier \etal\ 2021; Narayanan \etal\ 2021). 
 
In Figures A1 to A4 of Appendix A, we present for each ionic transition the observed absorption profile, accompanied by the best-fit Voigt profiles for individual components and the integrated Voigt profile for all components. We summarize the results of our Voigt profile analysis in Tables A1 to A4 of Appendix A for each of the four LLSs, where we include the best-fit Voigt profile parameters and their associated 68\% confidence intervals for each absorption component. In addition to the reported statistical errors, systematic uncertainty on the velocity centroid $dv_c$ is driven by the COS wavelength zero-point error, which is estimated to be $\lesssim3\,\kms$ based on the excellent agreement found between low-ionization absorption features in COS and optical Echelle spectra. Systematic uncertainty on absorption column densities is primarily due to continuum placement error, which may introduce up to $\approx0.05$ dex to the error budget of log\,$N_c$, for typical $S/N$ in our data. When all available transitions for a given species are saturated, we report instead the 95\% lower limits on log\, $N_c$ and the corresponding 95\% upper limits on $b_c$. Conversely, when a given species is not detected in all available transitions, we report the 95\% upper limits on $N_c$, calculated using the error spectrum over a spectral window that is twice the spectral FWHM. This calculation assumes an optically thin absorption line with $b_c=10\, \kms$ for low- and intermediate-ionization species or $b_c=30\, \kms$ for high-ionization \ion{O}{VI} and \ion{N}{V}.

\subsection{Ionization Modeling}

\begin{figure}
\includegraphics[scale=1.19]{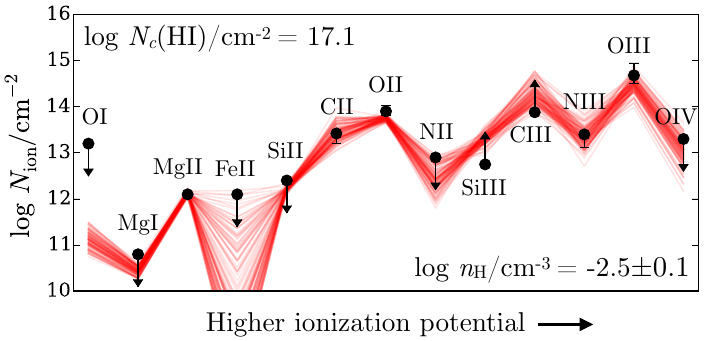}
\vspace{-0.5em}
\caption{Example of ionization analysis results for an optically thick component in the CUBS sample, in the LLS at $z_\mathrm{abs}=0.43526$ toward J0357$-$4812. The black circles indicate the observational constraints for different ionic column densities, ordered by increasing ionization energy, based on our Voigt profile analysis (\S 3.1). The ionization energies probed by the ionic species shown here range from $\approx0.6$ Ryd (\ion{Mg}{II}) to $\approx 4$ Ryd (\ion{O}{IV}). The thin red curves show 100 random MCMC realizations of the predicted ionic column densities from \textsc{cloudy} (\S 3.2). For a majority of individual absorbing components in the CUBS LLS sample, the observed ionic column densities are well-reproduced by a single photoionized gas phase.
}
\label{figure:ions}
\end{figure}

Constraining the detailed physical properties of the LLSs requires some knowledge of the gas ionization state. 
The cool gas temperatures inferred from the observed linewidths of different transitions with kinematically matched components ($T\sim10^4$ K, see \S 4.1) indicate that photoionization is the dominant ionization mechanism. For a photoionized gas, the ionization state and physical conditions can be inferred by comparing the observed column densities of different ions to their predicted values from different photoionization models. In particular, the ionization state is described by the dimensionless ionization parameter $U$, which is the number of incident ionizing photons per hydrogen 
atom. For a given radiation field, $U$ is inversely proportional to the hydrogen number density $n_\mathrm{H}$ according to $U\equiv \Phi/(c\,n_{\rm H})$, 
where $\Phi$ is the total flux of hydrogen-ionizing ($\geq1$ Ryd) photons.  In addition, an important physical quantity that affects the ionization state is the gas metallicity: a metal-enriched gas cools more efficiently than a low-metallicity one, so the photoionization equilibrium shifts toward lower ionization states. 

To gauge the gas ionization state, we perform a series of calculations using the \textsc{Cloudy} v.13.03 (Ferland \etal\ 2013) photoionization package. Our calculations assume a plane-parallel gas column with a uniform density $n_\mathrm{H}$, which is exposed to an ultraviolet background (UVB) radiation field that is normalized at $z=0.5$, roughly the median redshift of the CUBS LLSs. We consider a number of UVB models which exist in the literature. These UVBs include an updated version of the Haardt \& Madau (2001) UVB (known as HM05 in \textsc{Cloudy}), a more recent version of the Haardt \& Madau UVB (HM12 in \textsc{Cloudy}; Haardt \& Madau 2012), the UVB model by Khaire \& Srianand (2019; KS19 in in \textsc{Cloudy}), and a recent UVB model by Faucher-Gigu{\`e}re (2020; FG20). These different UVBs span a variety of UV slopes and total ionizing fluxes. The largest difference is between HM05 and HM12:  although HM05 is significantly softer than HM12 at energies between 1 and 10 Ryd, it has significantly more low-energy ($<3$ Ryd) photons and an overall higher number of ionizing photons (0.4 dex) compared to HM12 (e.g., Zahedy \etal\ 2019a). In addition, the HM12 UVB overpredicts the amplitude of the low-redshift ($z<1$) \ion{H}{I} column density distribution function by a factor of $\approx2$ (Shull \etal\ 2015). This discrepancy is a consequence of the assumption of a negligible escape fraction of hydrogen-ionizing photons from low-redshift galaxies in HM12 (e.g., Shull \etal\ 2015), leading to a low inferred \ion{H}{I} photoionization rate ($\Gamma_\mathrm{\ion{H}{I}}$). Subsequent studies have found that UVBs with inferred $\Gamma_\mathrm{\ion{H}{I}}$ that are $\approx1.5-3$ times higher than that of the HM12 UVB provide better agreement with low-redshift observations (e.g., Khaire \& Srianand 2015; Viel \etal\ 2017). 

In contrast, the FG20 and KS19 UVB models are informed by the latest observations of the low-redshift Ly$\alpha$ forest and provide a good match to empirical constraints on $\Gamma_\mathrm{\ion{H}{I}}$ at $z<0.5$ (e.g., Gaikwad \etal\ 2017; Khaire \etal\ 2019). While FG20 and KS19 exhibit significant differences in the less well-constrained $\sim10-100$ Ryd (soft X-ray) regime, the two UVBs agree well at $\lesssim 5$ Ryd. Similarly, FG20 has a nearly identical slope to the HM05 UVB at energies $\lesssim 4$ Ryd which are needed to create most of the ionic species encountered in our analysis. However, its overall flux of ionizing photons is $\approx 0.2$ dex ($\approx60$\%) lower than HM05 at $z\approx0.5$. As discussed in \S 4.2, we find little systematic differences ($\lesssim0.1$ dex) in inferred metallicities using FG20 compared to HM05, smaller than the typical statistical uncertainties of these quantities in our analysis. However, the inferred gas densities (cloud sizes) are up to $\approx0.2$ dex lower (larger) using FG20 compared to HM05, which is explained by FG20's lower flux of ionizing photons. 

We choose the HM05 UVB as the fiducial ionizing radiation field in our study in order to facilitate comparisons with recent photoionization studies of the $z<1$ CGM (e.g., Zahedy \etal\ 2019a; Lehner \etal\ 2019). 
However, wherever appropriate throughout the paper, we highlight the similarities/differences between HM05 and FG20 and how they impact our conclusions. As a reference, the relationship between $U$ and $n_\mathrm{H}$ is described by log\,$U = -5.31  - \mathrm{log}\, n_\mathrm{H}/\cmjjj$ for the HM05 UVB at $z=0.5$. For FG20, it is log\,$U = -5.52  - \mathrm{log}\, n_\mathrm{H}/\cmjjj$.
We also note that while we assume a single UVB redshift of $z=0.5$ in all of our calculations, changing the UVB redshift to either the minimum ($z=0.36$) or maximum  ($z=0.62$) redshift of the absorbers would change the resulting UVB intensity by no more than $\approx0.1$ dex, with a negligible change to the UVB slope. The inferred metallicity would thus be unchanged, while the inferred gas density and cloud size would change by $\approx\pm0.1$ dex. However, this effect is subdominant compared to the systematic error on inferred gas density and metallicity due to the uncertain UVB slope ($0.3-0.4$ dex; e.g., Zahedy \etal\ 2019), as well as the uncertainties on recent $\Gamma_\mathrm{\ion{H}{I}}$ measurements at $z<1$  ($\approx 0.2$ dex; e.g. Gaikwad \etal\ 2017; Khaire et al. 2019).

To infer the physical conditions of the gas, we construct a grid of \textsc{Cloudy} models that represent a diversity of physical characteristics. These models span a wide range of \ion{H}{I} column densities ($14\leq \mathrm{log}\,N\mathrm{(\ion{H}{I})/cm^{-2}}\leq 20$ in 0.25 dex steps), number densities ($-5\leq \mathrm{log}\,n\mathrm{_H/cm^{-3}}\leq 1$ in 0.25 dex steps), 
and  of $\alpha$-element abundances ($-3\leq\mathrm{[\alpha/H]}\leq1$ in 0.25 dex steps). We choose $\alpha$ elements as tracers of the gas metallicity due to the relatively large number of available O, Mg, Si, and S ionic species in our dataset (\S 3.1).  For each grid point, \textsc{Cloudy} calculates the expected ionic metal
column densities and their ionization fractions assuming photoionization equilibrium. To explore the possibility of a non-solar chemical abundance pattern of the gas, we allow the gas to depart from solar values in $\mathrm{[C/\alpha]}$, $\mathrm{[N/\alpha]}$, and $\mathrm{[Fe/\alpha]}$ relative abundance ratios. Because C, N, and Fe are produced in various nucleosynthetic pathways over different timescales (e.g., Chiappini \etal\ 2003; Kobayashi \etal\ 2020b), investigating their abundances relative to $\alpha$ elements (which are produced primarily in massive stars) may offer valuable insight into the chemical enrichment history of the gas. 

We compare the resulting \textsc{Cloudy} predictions to the data by performing
a statistical analysis that considers not only the different measured ionic column densities but also upper limits (non-detections) and lower limits (saturation). Given a set of gas density, metallicity, and chemical abundance pattern, we define the likelihood of observing a set of kinematically well-matched ionic species $\{y_i\}$ with $n$ column density measurements, $m$ upper limits, and $l$ lower limits to be 

{\small
\begin{align}
{\cal P} \propto  \left( \prod_{i=1}^{n} \exp \left\{ -\frac{1}{2} \left[ \frac{y_i -
\bar{y_i}(n_\mathrm{H},\mathrm{[\alpha/H]},\mathrm{[X/\alpha]})}{\sigma_i} \right]^2 \right\} \right)  \nonumber 
\\  
\times\left(\prod_{i=1}^{m} \int_{-\infty}^{y_i} dy' \exp \left\{ -\frac{1}{2} \left[ 
\frac{y' - \bar{y_i}(n_\mathrm{H},\mathrm{[\alpha/H]},\mathrm{[X/\alpha]})}{\sigma_i} \right]^2 \right\} \right)  \nonumber
\\
\times\left(\prod_{i=1}^{l} \int_{y_i}^{+\infty} dy' \exp \left\{ -\frac{1}{2} \left[ 
\frac{y' - \bar{y_i}(n_\mathrm{H},\mathrm{[\alpha/H]},\mathrm{[X/\alpha]})}{\sigma_i} \right]^2 \right\} \right),
\end{align}}
where X=C, N, and Fe, $y_i=\log\,N_i$ is the column density constraint of the $i$-th species and $\sigma_i$ is its estimated uncertainty,
and $\bar{y_i}=\log\,\bar{N_i}$ is the model prediction. While the first product in Eq. 1 is equivalent to calculating $e^{-\frac{1}{2}\chi^{2}}$ for the measured ionic column densities, the second and third products extend the calculation to non detections and saturated lines, respectively (see also Zahedy \etal\ 2019a). 

\begin{figure}
\includegraphics[scale=0.68]{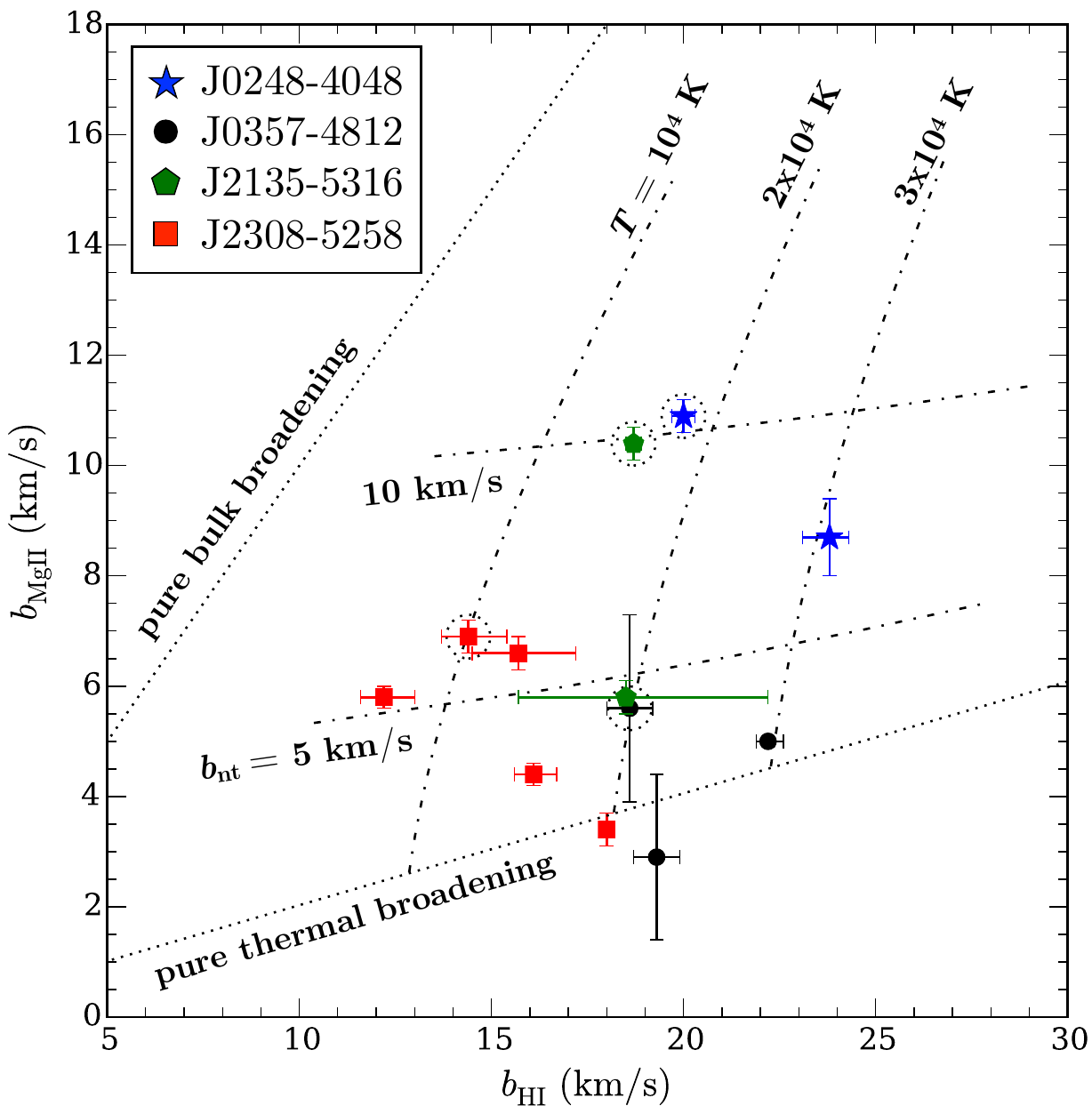}
\vspace{-0.5em}
\caption{Observed distribution of linewidths for matched \ion{Mg}{II} and \ion{H}{I} components. These components have some of the higher $N_c\mathrm{(\ion{H}{I})}$ in the sample, log\,$N_c\mathrm{(\ion{H}{I})}/$\cmjj$ \gtrsim16$. Optically thick components (log\,$N_c\mathrm{(\ion{H}{I})}/$\cmjj$ \gtrsim17.2$) are marked with dotted circles. 
Two limiting cases are indicated:  the lower dotted line indicates purely thermally broadened gas, where $b_\mathrm{\ion{Mg}{II}}\approx0.2\,b_\mathrm{\ion{H}{I}}$, while the upper dotted line shows a pure bulk broadening case with $b_\mathrm{\ion{Mg}{II}}\approx b_\mathrm{\ion{H}{I}}$. The dash-dotted curves indicate various combinations of gas temperatures and non-thermal broadening.  \ion{Mg}{II}-bearing cool clouds in the CUBS LLS sample have a mean kinetic temperature and dispersion of $\langle T \rangle =2\times10^4\,$K and $\sigma_T =1\times10^4\,$K, with a mean non-thermal/bulk broadening of $\langle b_\mathrm{nt} \rangle =5\pm3\,$\kms.}
\label{figure:ions}
\end{figure}

\begin{figure*}
\includegraphics[scale=0.62]{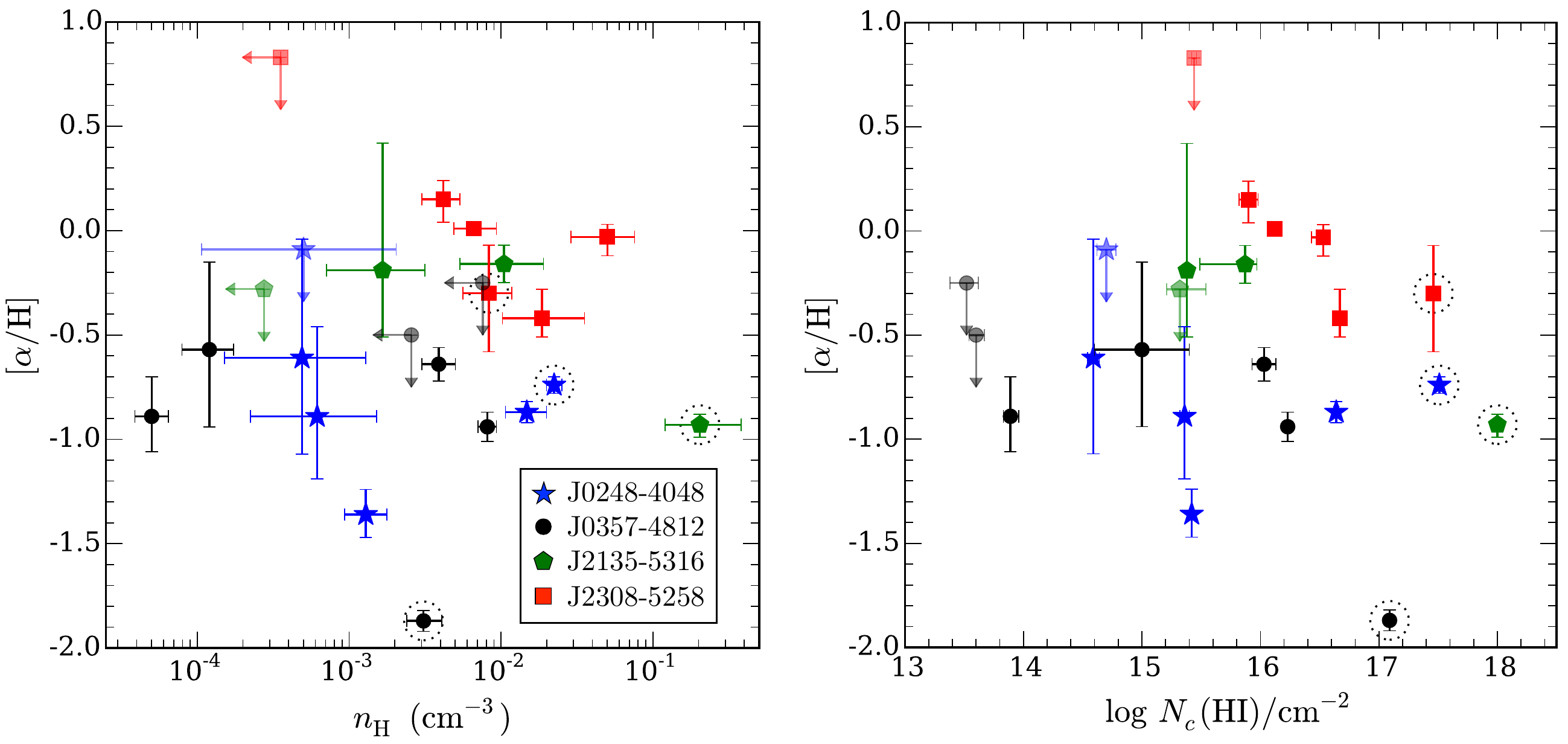}
\vspace{-0.5em}
\caption{{\it Left:} $\mathrm{[\alpha/H]}$ versus $n_\mathrm{H}$ for individual components in the CUBS LLS sample. For each data point, the vertical and horizontal error bars indicate the 68 percent confidence intervals for $\mathrm{[\alpha/H]}$ and $n_\mathrm{H}$, respectively. Optically thick components are marked with dotted circles. When only upper/lower limits are available for a given component, it is presented as a faded data point with arrows showing the 95 percent upper/lower limits on $\mathrm{[\alpha/H]}$ and/or $n_\mathrm{H}$. There is no significant correlation between $\mathrm{[\alpha/H]}$ and $n_\mathrm{H}$. {\it Right}: $\mathrm{[\alpha/H]}$ versus $N_c\mathrm{(\ion{H}{I})}$. The estimated median metallicity is $\mathrm{[\alpha/H]_{1/2}}=-0.7^{+0.1}_{-0.2}$ among individual components, with a 16-84 percentile range of $\mathrm{[\alpha/H]}=(-1.3,-0.1)$.}
\label{figure:ions}
\end{figure*}

The statistical analysis described above is performed for each absorbing component previously identified in \S3.1. 
We adopt an MCMC approach using the \textsc{emcee} package to compare the observational constraints to an interpolated grid of \textsc{Cloudy} models (evaluated at the observed $N\mathrm{(\ion{H}{I})}$) and construct a posterior PDF for each physical parameter. 
Each MCMC run consists of 500 steps performed by a group of 500 walkers, with a ``burn-in" of 100 steps per chain. To speed up convergence, the walkers  are seeded in a small, high probability region within the parameter space. These initial guesses are determined via a ``by-eye" visual comparison between the observed ionic column densities and the corresponding model predictions, evaluated in a coarse grid that spans the full parameter space.   

By default, we exclude measurements of high-ionization species \ion{O}{VI} from the statistical analysis above, due to the uncertainty in the ionization mechanism of \ion{O}{VI}-bearing gas (e.g., Oppenheimer \etal\ 2016; 2018; Werk \etal\ 2016; Stern \etal\ 2018) and the frequently observed kinematic misalignments between \ion{O}{VI} absorption profiles and those of lower-ionization species (e.g., Savage \etal\ 2010; Zahedy \etal\ 2019a; 2020; Rudie \etal\ 2019). Such kinematic misalignments suggest that \ion{O}{VI}-bearing gas resides in a more highly ionized (and likely higher temperature) phase which is distinct from the lower-ionization, photoionized gas probed by \ion{H}{I} and other low ions (e.g., Zahedy \etal\ 2019a; 2020). 

In a minority of cases (four components in two LLSs or $\approx30$ percent of \ion{O}{VI} components identified), the presence of narrow \ion{O}{VI} components ($b_c\lesssim20$ \kms) that are kinematically well aligned with lower-ionization species (e.g., \ion{H}{I}, \ion{C}{III}, and \ion{O}{IV}) suggests the possibility of photoionized \ion{O}{VI}-bearing gas and justifies their inclusion in our ionization analysis. In these few instances, we find that the photoionization model is able to achieve a good fit to the data (see \S 5.1 and discussion on individual systems in Appendix A), with inferred low gas densities of $n_\mathrm{H}\sim10^{-4}\, \cmjjj$. 
In general, our model calculations indicate that a higher ionization gas phase traced by  \ion{O}{VI} contributes negligibly  to the observed column densities of low ions. For that reason, the default exclusion of \ion{O}{VI} from our ionization modeling should not result in a significant bias in the inferred $U$ values of the lower-ionization gas phase. 

For the most part, the observed ionic column densities of various low- and intermediate-ionization metal species are well reproduced by a single photoionized gas phase, an example of which is shown in Figure 1. One exception is the component at $dv_c=+32$ \kms\ in the LLS at $z_\mathrm{abs}=0.43526$ toward sightline J0357$-$4812. As discussed in \S 5.1 and Appendix A2, a single-phase model fails to reproduce the observed \ion{O}{IV} column density, under-predicting it by more than an order of magnitude. In contrast, a two-phase gas composed of a high-density phase with  $n_\mathrm{H}\approx4\times10^{-3}$ \cmjjj\ and a low-density phase with $n_\mathrm{H}\approx10^{-4}$ \cmjjj\ is required to reproduce all the observed ionic column densities. A more detailed discussion on the multiphase nature of CGM absorbers associated with LLSs is provided in \S 5.1. 
  
We present the results of the ionization analysis in Table A5 in Appendix A, where for each component we report 
the most probable $\mathrm{[\alpha/H]}$, $n_\mathrm{H}$, and elemental abundance ratios $\mathrm{[N/\alpha]}$, $\mathrm{[Fe/\alpha]}$, and $\mathrm{[C/\alpha]}$, as well as the estimated 68\% confidence interval for each quantity. For components with two or fewer associated metal-line detections, which holds for $\approx25$ percent of all components, the inferred values are subject to large uncertainties ($\gtrsim0.5$ dex), and in some cases there is no clear probability maximum within the boundaries of the parameter space. For such components, we report instead the estimated 95\% upper/lower limits for the different parameters in Table A5. 

\begin{figure*}
\includegraphics[scale=0.65]{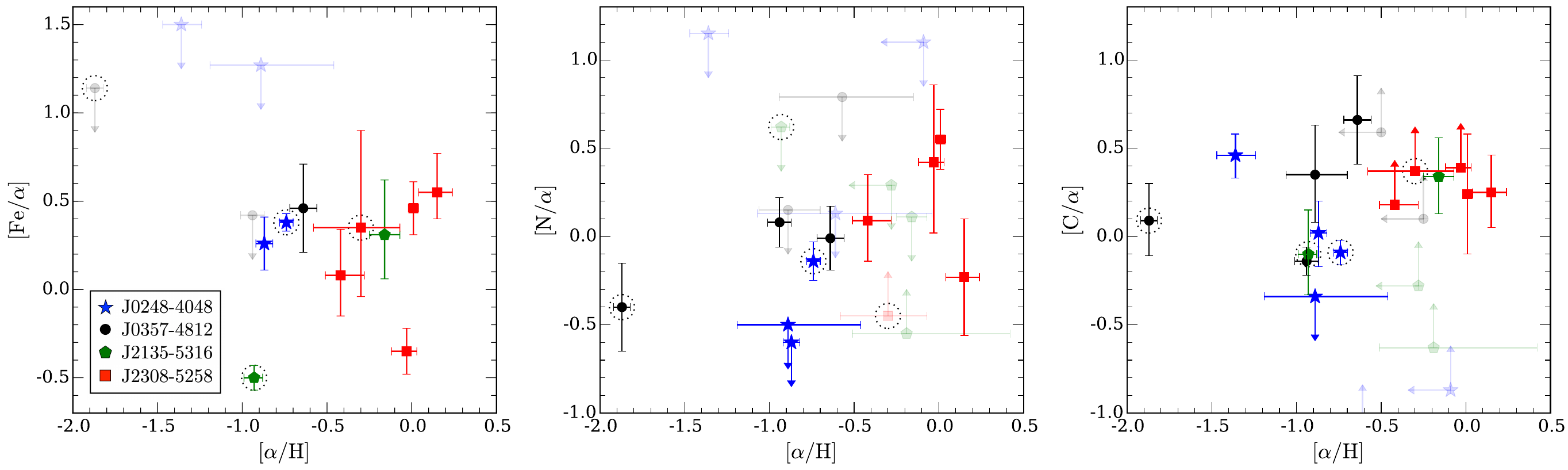}
\vspace{-0.5em}
\caption{
$\mathrm{[Fe/\alpha]}$, $\mathrm{[N/\alpha]}$,  and $\mathrm{[C/\alpha]}$ versus metallicity in CUBS LLS sample. In each panel, optically thick components are marked with dotted circles. Components with poorly constrained elemental abundance ratios have been faded out for clarity. These absorbers exhibit a wide range of inferred elemental abundance ratios, which indicate a diversity of chemical enrichment histories.}
\label{figure:ions}
\end{figure*}

\section{Results}

We present a discussion on the gas properties of individual LLSs in Appendix A, which we summarize here. Our absorption analysis and ionization modeling reveal a diversity of gas properties in the four LLSs. All four absorbers exhibit a multicomponent kinematic structure, seen in \ion{H}{I} and various associated metal lines, that is spread over several hundred \kms\ in velocity space. The excellent kinematic agreements between \ion{H}{I}, low ions, and intermediate ions indicate that these different species are physically associated and justify modeling them as a single gas phase (with some exceptions described in \S 5.1). Here we discuss the physical properties and elemental abundances of these four $z<1$ LLSs. 

\subsection{Thermal properties}

In addition to allowing an ionization analysis to infer gas density and elemental abudances, our detailed absorption analysis also constrains the gas temperature and non-thermal motion within cool clumps associated with LLSs. Recent cosmological simulations show that a significant fraction of strong absorbers with  log\,$N$(H\,I)/\cmjj$ \gtrsim17$ trace accreting gas in the CGM at low and high redshifts (e.g., van de Voort \etal\ 2012; Faucher-Gigu{\`e}re \etal\ 2015; Hafen \etal\ 2017). Investigating the thermal properties of LLSs  may provide a valuable testing ground for this theoretical picture. 

We find that \ion{H}{I}-bearing gas in the CUBS LLS sample is predominantly cool, with most components consistent with having temperatures of $T\lesssim(1-2)\times 10^5$ K, inferred from the observed \ion{H}{I} linewidths of $b_c\mathrm{(\ion{H}{I})}\lesssim60$ \kms\ assuming purely thermal line broadening (Table A1 to Table A4; see also Tumlinson \etal\ 2013). To obtain a more precise constraint on gas temperature, it is necessary to compare the linewidths of different absorption species. Because line broadening arises from both thermal and non-thermal processes (e.g., turbulence and bulk motions), it is possible to determine the relative contributions of the two processes when transitions of significantly different atomic weights are observed simultaneously. 

In Figure 2, we show the observed linewidths for matched \ion{Mg}{II} and \ion{H}{I} absorption components in the sample. These components have some of the higher \ion{H}{I} column densities in the sample, log\,$N$(HI)/\cmjj$ \gtrsim16$. With the exception of a couple of broad components, most \ion{Mg}{II} components are narrow, with $b_c\mathrm{(\ion{Mg}{II})}\lesssim6$ \kms, indicating that the gas is cool ($T\sim 10^4$ K). Furthermore, the lack of a trend in $b_c\mathrm{(\ion{Mg}{II})}$ versus $N$(\ion{H}{I}) indicates a similar thermal properties between optically thin and thick absorbers in the sample. While two optically thick components have the broadest \ion{Mg}{II} linewidths  $b_c\mathrm{(\ion{Mg}{II})}\approx10$ \kms\ in the sample, it may be explained by the presence of additional kinematic structures at a scale unresolved by our data. 

Also drawn in Figure 2 are two dotted lines which indicate limiting cases for the line broadening mechanism. The dotted line in the bottom indicates a pure thermal broadening case. For a thermally broadened gas, the \ion{Mg}{II} and \ion{H}{I} linewidths are related by the square root of the ratio of their atomic masses alone, or $b_c\mathrm{(\ion{Mg}{II})}\approx0.2\,b_c\mathrm{(\ion{H}{I})}$. In contrast, a second dotted line shows the limiting case of purely non-thermal line broadening, where $b_c\mathrm{(\ion{Mg}{II})}\approx b_c\mathrm{(\ion{H}{I})}$. The majority of data points are located closer to the pure thermal-broadening line than to the pure non-thermal broadening line. Together with the narrow observed Mg\,II linewidths, the observed $b_c\mathrm{(\ion{Mg}{II})}/b_c\mathrm{(\ion{H}{I})}$ ratios indicate that gas temperature is the primary broadening mechanism. Specifically, gas associated with matched \ion{H}{I} and \ion{Mg}{II} kinematics components in the sample has a mean temperature and dispersion of $\langle T \rangle =2.0\times10^4\,$K and $\sigma_T =1\times10^4\,$K, with a relatively modest non-thermal/bulk line broadening of $\langle b_\mathrm{nt} \rangle =5\pm3\,$\kms. 
While we do not consider gas temperature as a free parameter in our ionization modeling, the measured temperatures of these \ion{Mg}{II}-bearing components are broadly consistent with the output temperatures from the best-fit \textsc{Cloudy} photoionization models. Furthermore, the observed temperatures and non-thermal line broadening in our sample are similar to what have been reported in the gaseous haloes of massive ellipticals (Zahedy \etal\ 2019a) and indicate that these gas clouds could arise in dynamically cool structures in the CGM.

\subsection {Metallicity and elemental abundance ratios}

Our ionization analysis based on matched kinematic  components has revealed significant variations in chemical abundances and gas densities in absorbers associated with $z<1$ LLSs. The left panel of Figure 3 shows the component $\mathrm{[\alpha/H]}$ versus $n_\mathrm{H}$ in the sample and illustrates the diversity of observed gas metallicities and densities. While there is no statistically significant signature of a correlation between $\mathrm{[\alpha/H]}$ and $n_\mathrm{H}$, these absorbers occupy a wide range of metallicities ($\approx0.01$ solar to $\approx$solar) and densities (from $n_\mathrm{H}\sim10^{-4}\, \cmjjj$ to $n_\mathrm{H}\sim0.1\, \cmjjj$). 

The left panel of Figure 3 reveals another interesting feature. We find significant variations in $\mathrm{[\alpha/H]}$ among different components within individual LLS. Specifically, all four LLSs exhibit a metallicity difference of a factor of $\approx5$ or larger between the highest- and lowest-metallicity components. Large variations in gas metallicities within the gaseous haloes of individual galaxies have been reported around $z\approx0.5$ massive quiescent galaxies (Zahedy \etal\ 2019a) and are likely attributable to poor chemical mixing in the CGM (see also D'Odorico \& Petitjean 2001). 
In addition, metal-rich components with $\mathrm{[\alpha/H]}\gtrsim-1.0$ are associated with all four LLSs in the sample, whereas metal-poor components with $\mathrm{[\alpha/H]}\lesssim-1.0$ are associated with two out of four systems. Across the CUBS LLS sample, the median metallicity of individual kinematic components is $\mathrm{[\alpha/H]_{1/2}}=-0.7^{+0.1}_{-0.2}$. The uncertainty in the median gas metallicity is estimated using a combined bootstrap and Monte Carlo resampling method. In addition, the estimated 16-84 percentile range in gas metallicity is  $\mathrm{[\alpha/H]}=(-1.3,-0.1)$. Note that we have excluded components with poorly constrained $\mathrm{[\alpha/H]}$ from these calculations, i.e., those with metallicity upper limits that are higher than solar metallicity. 

The distribution of inferred gas metallicities in the CUBS LLS sample is in agreement with the observed chemical enrichment level in the CGM of massive quiescent galaxies (Zahedy \etal\ 2019a), and with the metallicity distribution of $z<1$ LLSs in a number of recent high-resolution cosmological simulations. In particular, the FIRE simulation project reported a metallicity distribution consistent with a single mode characterized by a median metallicity and dispersion of $\mathrm{[M/H]_{1/2}}=-0.9\pm0.4$ (Hafen \etal\ 2017), which is similar to the median LLS metallicity of $\mathrm{[M/H]_{1/2}}=-0.8$ reported in the EAGLE simulations at $z\approx0.5$ (Rahmati \& Oppenheimer 2018).\footnote{These theoretical studies define the metallicity $\mathrm{[M/H]}$ as the total abundance of heavy elements relative to solar. Because $\alpha$ elements like O, Mg, and Si comprise the majority of heavy elements both in mass and abundance, the metallicities from these studies can be straightforwardly compared with our measurements.} The median metallicity in our sample is also roughly consistent with recent measurements of $z<1$ LLSs (Lehner \etal\ 2019), although these authors reported a higher fraction ($\approx$10 \%) of very metal-poor absorber with $\mathrm{[M/H]}\lesssim-2$. 

\begin{figure}
\hspace{-1em}
\includegraphics[scale=0.65]{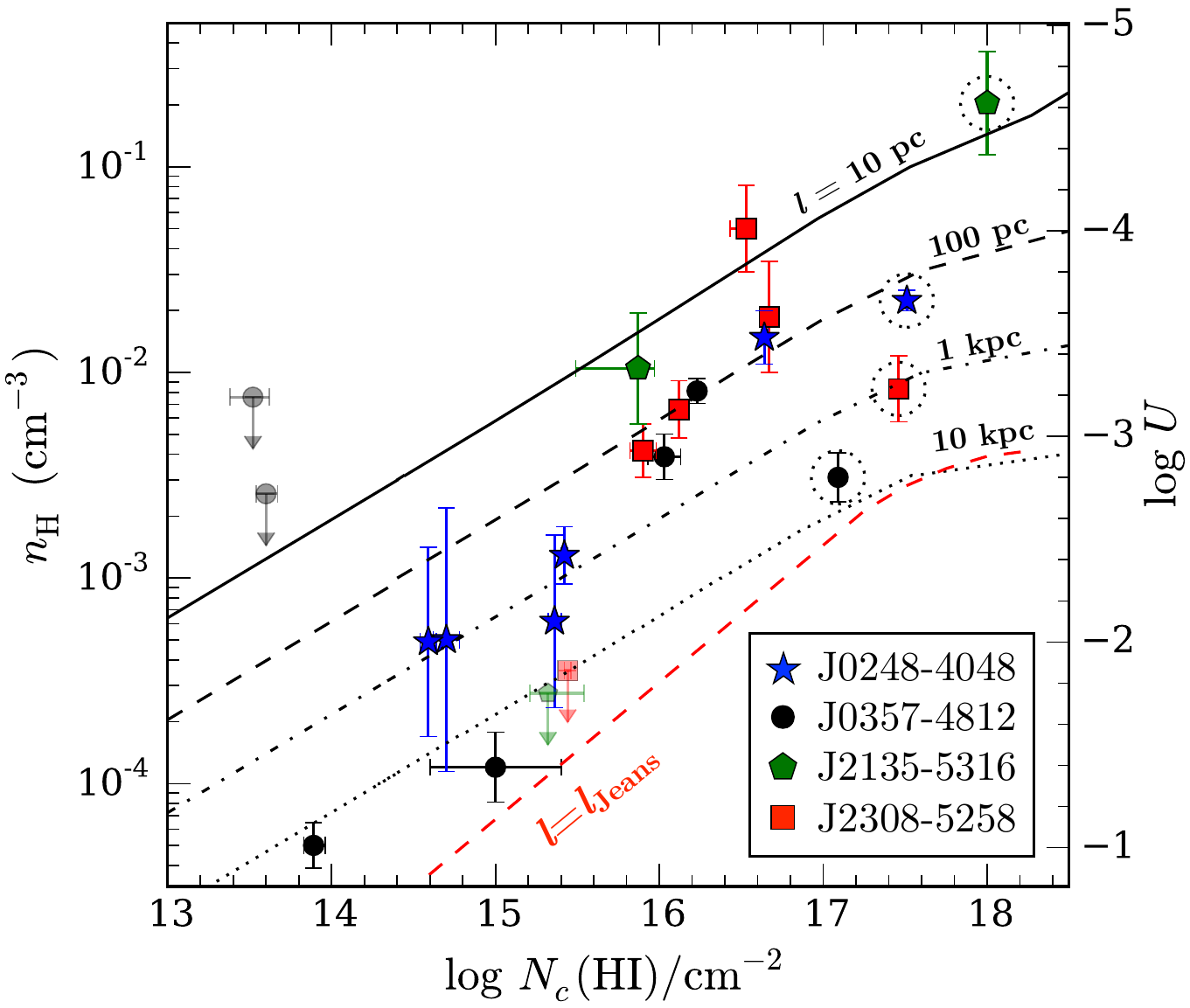}
\vspace{-2.em}
\caption{
Gas number density $n_\mathrm{H}$ plotted versus $N_c\mathrm{(\ion{H}{I})}$. For each  $n_\mathrm{H}$, the corresponding ionization parameter $U$ is also shown assuming the HM05 UVB. The data exhibit a trend of increasing $n_\mathrm{H}$ with $N_c\mathrm{(\ion{H}{I})}$:  higher-$N_c\mathrm{(\ion{H}{I})}$ components are less ionized than more optically thin components. The median $U$ in CUBS LLS sample is $\mathrm{log}\,U_{1/2}\approx-2.9$. In addition, each black-colored curve shows the expected relationship between $n_\mathrm{H}$ and $N_c\mathrm{(\ion{H}{I})}$ for a cloud with a given thickness, from $\ell=10$ pc to $\ell=10$ kpc. For comparison, the red dashed curve shows the expected relation for self-gravitating gas clouds (Schaye 2001).
}
\label{figure:ions}
\end{figure}

Next, we show $\mathrm{[\alpha/H]}$ versus $N_c\mathrm{(\ion{H}{I})}$ in the right panel of Figure 3. There is no statistically significant trend in $\mathrm{[\alpha/H]}$ versus $N_c\mathrm{(\ion{H}{I})}$, which is contrary to a number of recent studies reporting an anti correlation between gas metallicity and $N\mathrm{(\ion{H}{I})}$ in the $z<1$ CGM (e.g., Prochaska \etal\ 2017). 
However, it should be noted that these studies employed the HM12 UVB, which is known to systematically predict progressively higher gas metallicities (up to $\approx 0.7$ dex) with decreasing $N\mathrm{(\ion{H}{I})}$ compared to HM05 (e.g., Chen \etal\ 2017; Zahedy \etal\ 2019a; Wotta \etal\ 2019). Because the HM12 UVB is significantly harder than HM05, as $N\mathrm{(\ion{H}{I})}$ decreases and the gas becomes more ionized, more low- and intermediate-ionization species (e.g., \ion{Mg}{II}, \ion{C}{II}, \ion{Si}{III}) are  preferentially lost to higher ionization species in HM12. This systematic effect can fully explain the reported anti correlation between gas metallicity and $N$(\ion{H}{I}) (see Zahedy \etal\ 2019a), because metallicity inferences frequently rely on observing various low- and intermediate-ionization metal species.  In contrast, we find no significant systematic differences ($\lesssim0.1$ dex) in inferred gas metallicities using FG20 compared to HM05, owing to their very similar slopes at $\lesssim4$ Ryd. 

Furthermore, the high quality of the QSO spectra in CUBS allows us to investigate the detailed elemental abundance ratios in the LLS sample. Recall that in our ionization analysis, we consider the elemental abundance ratios of $\mathrm{[C/\alpha]}$, $\mathrm{[N/\alpha]}$, and $\mathrm{[Fe/\alpha]}$ as free parameters in the model.  We are able to obtain precise constraints on the gas-phase elemental abundance ratios for approximately 50 percent of the components we have identified. The observed relative abundances of different elements represent a fossil record of different sources of heavy element production. Because each of these elements is produced over different timescales in stars of various masses (see discussion in \S\ 5.2), their relative abundances reflect the chemical enrichment history and offer critical insight into the physical origin of the gas (e.g., Zahedy \etal\ 2016; 2017). 

\begin{figure*}
\includegraphics[scale=1.1]{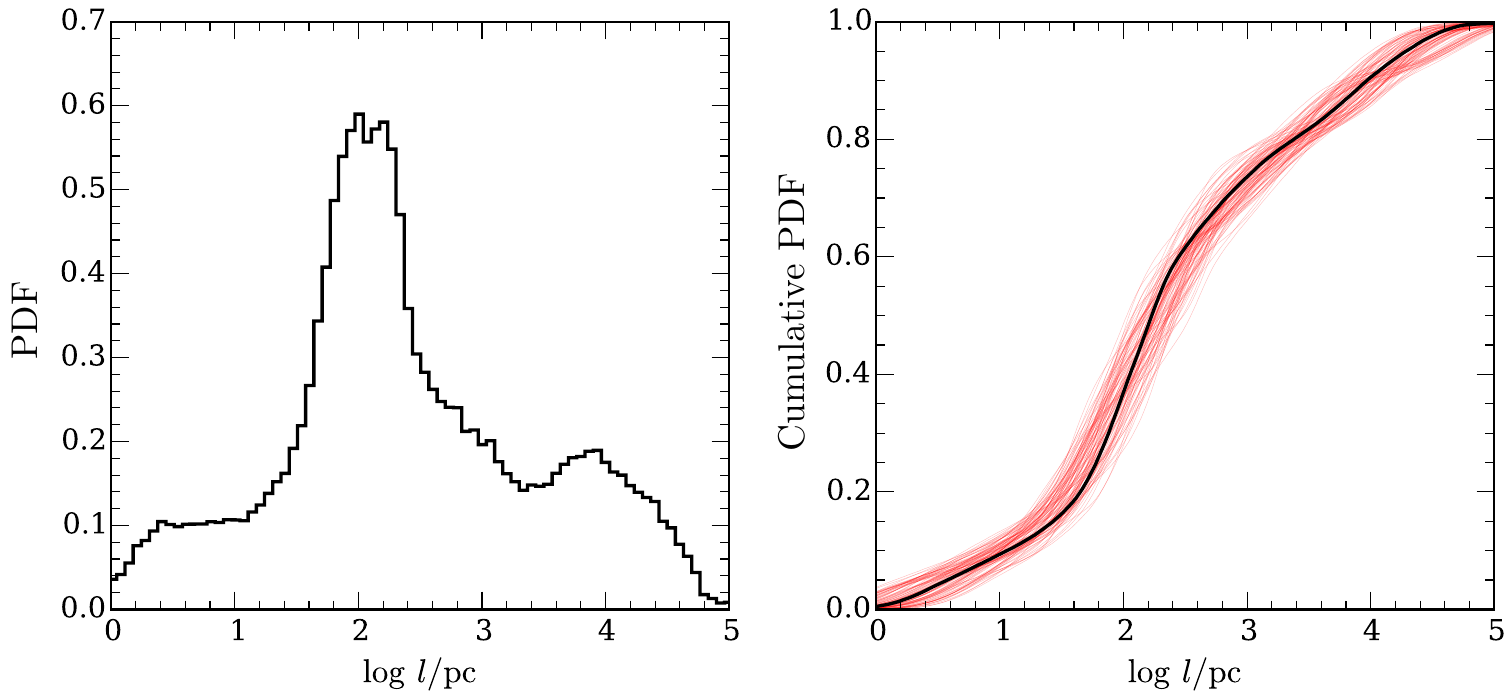}
\vspace{-0.5em}
\caption{
 {\it Left}: PDF of inferred cloud thickness for individual absorption components in the CUBS LLS sample. We estimate a median cloud size of $l_\mathrm{1/2} =160^{+140}_{-50}$  pc in the CUBS LLS sample using the HM05 UVB. {\it Right}: The corresponding cumulative PDF of cloud thickness (thick black curve). The thin red curves demonstrate how uncertainties in the UVB intensity would affect the inferred cloud thickness. These curves are constructed from 100 random resamplings of the cloud size PDF by the Monte Carlo method, obtained by allowing the UVB intensity to randomly vary in individual components by up to $\pm0.5$ dex.}
\label{figure:ions}
\end{figure*}

In the three different panels of Figure 4, we present the relative abundance ratio of $\mathrm{[Fe/\alpha]}$, $\mathrm{[N/\alpha]}$, and $\mathrm{[C/\alpha]}$ versus $\mathrm{[\alpha/H]}$ for individual components in the CUBS LLS sample. There are a number of features that are worth highlighting. 
First, our analysis reveals a wide range of elemental abundance ratios within and among the four LLSs studied, which indicates a variety of chemical enrichment histories of the gas. Despite these variations, there is hint of a trend of  rising $\mathrm{[N/\alpha]}$, $\mathrm{[C/\alpha]}$, and $\mathrm{[Fe/\alpha]}$ abundance ratios with increasing gas metallicity, although it is not statistically significant. Given the relatively small size of the current sample, more data are needed to confirm the existence of this trend. However, it is worth noting that trends of increasing $\mathrm{[N/\alpha]}$ and $\mathrm{[C/\alpha]}$ with gas-phase metallicity are well-known in the ISM of both normal star-forming and dwarf galaxies in the local Universe (e.g., van Zee \etal\ 2006; Esteban \etal\ 2009; 2014; Pilyugin \etal\ 2014; 2015; Berg \etal\ 2016), and a trend of increasing $\mathrm{[Fe/\alpha]}$ with stellar metallicity ($\mathrm{[Fe/H]}$) is seen in Galactic disk and halo stars (e.g., McWilliam 1997; Akerman \etal\ 2004; Fabbian \etal\ 2009).

While there is often a large spread in chemical abundance ratios, in some cases different components within an absorption system share similar elemental abundance ratios. For instance, four components in the LLS toward J2308$-$5258 (red squares) exhibit $\mathrm{[Fe/\alpha]}\approx0.1-0.5$. Such similarities in elemental abundance ratios indicate the different gas clumps in the absorber likely share the same chemical enrichment history, even if there are still differences in the overall gas metallicity. In \S 5.2, we explore the physical relationship between the observed elemental abundance ratios in these LLSs with the known elemental abundances in stellar populations and star-forming regions in galaxies. 

\subsection {Gas densities and cloud sizes}

Figure 5 demonstrates a wide range of gas densities in absorbers associated with $z<1$ LLSs, ranging from $n_\mathrm{H}\approx10^{-4}\, \cmjjj$ to $n_\mathrm{H}\approx0.1\, \cmjjj$. To investigate how gas density depends on $N_c\mathrm{(\ion{H}{I})}$, we plot $n_\mathrm{H}$ versus $N_c\mathrm{(\ion{H}{I})}$ in Figure 5. The data show a clear trend of increasing gas density with rising $N_c$(\ion{H}{I}), which indicates that more optically thick gas is less ionized than its optically thinner counterpart. Overall, the median $U$ in our sample is $\mathrm{log}\,U_{1/2}\approx-2.9$. In contrast, stronger absorption components with log\,$N_c\mathrm{(\ion{H}{I})/\cmjj}\gtrsim16$ have a median ionization parameter of $\mathrm{log}\, U_{1/2} \approx-3.3$. The inferred $U$ values for higher column density gas in our sample are in general agreement with what has been found in partial and full LLSs at $z<1$ (e.g., Lehner \etal\ 2013; 2019) and in absorbers with similar $N$(\ion{H}{I}) in the gaseous haloes of massive quiescent galaxies at $z\approx0.5$ (Zahedy \etal\ 2019a).

Furthermore, the observed correlation between $N_c\mathrm{(\ion{H}{I})}$ and $n_\mathrm{H}$ indicates a well-defined distribution of physical sizes for cool clumps associated with LLSs. In Figure 5, we plot the inferred $n_\mathrm{H}$ and $N_c\mathrm{(\ion{H}{I})}$ for individual components. We also present different curves showing the expected relationship between $n_\mathrm{H}$ and $N\mathrm{(\ion{H}{I})}$ for clouds with different line-of-sight thicknesses, assuming they are photoionized by the HM05 UVB. It is apparent that the majority of cool clumps are spatially compact, with thicknesses between $\approx10$ pc and $\approx10$ kpc and a mode of around $\sim100$ pc. For comparison, we indicate in Figure 5 the expected relationship between $n_\mathrm{H}$ and $N_c\mathrm{(\ion{H}{I})}$ for self-gravitating clouds with sizes $\ell=\ell_\mathrm{Jeans}$, where $\ell_\mathrm{Jeans}$ is the local Jeans length (Schaye 2001). If a cloud is significantly larger than its local Jeans length, it is prone to gravitational instability which may lead it to collapse and/or fragment. Based on this exercise, it is clear that gas clouds associated with the LLSs in our sample are smaller than their Jeans length, which indicate that they arise in stable structures that are likely to be in pressure equilibrium. 

In addition, there is also a hint of a correlation between cloud thickness and the ionization state of the gas. Specifically, a majority of components with log\,$N_c\mathrm{(\ion{H}{I})/\cmjj}\gtrsim16$ are less than a few hundred pc thick, whereas more optically thin (and highly ionized) clouds appear to be more spatially extended with thicknesses of  $\gtrsim1$ kpc. While the sample size is small and we caution against drawing firm conclusions at this point, this possible correlation is consistent with the physical picture in which more highly ionized gas in the CGM is more spatially extended and volume-filling than lower-ionization gas. 

To further examine the distribution of cloud sizes in the sample, we construct a PDF of cloud sizes by summing the posterior PDFs of all well-constrained individual absorption components. The cloud size PDF is plotted in the left panel of Figure 6. This figure shows a clear peak in the cloud-size PDF at a spatial scale of $\sim100$ pc. The median cloud thickness in our sample is $\ell_\mathrm{1/2} =160$ pc, which can be seen in the corresponding cumulative PDF shown in Figure 6. The estimated range of cloud thicknesses containing 68 percent probability for all individual components is $\ell=(25\, \mathrm{pc},4 \,\mathrm{kpc})$. In addition, a smaller peak in the PDF is seen on a scale of a few kpc, which can be attributed to optically thin gas with high ionization parameters. Several of these highly ionized components exhibit the presence of kinematically coincident $\ion{O}{VI}$ absorption which is well reproduced by our photoionization modeling (see discussion in \S 5.1). The range of inferred clump sizes in the CUBS LLS sample is in excellent agreement with previous findings for cool clouds in the Galactic halo (e.g., Hsu \etal\ 2011) and in the CGM of massive elliptical galaxies, where Zahedy \etal\ (2019) found a median size of  $\ell_\mathrm{1/2} \approx120^{+80}_{-40}$ pc and a 16-84 percentile range of $\ell=(20, 800)$ pc. 

\begin{figure}
\hspace{-1em}
\includegraphics[scale=0.87]{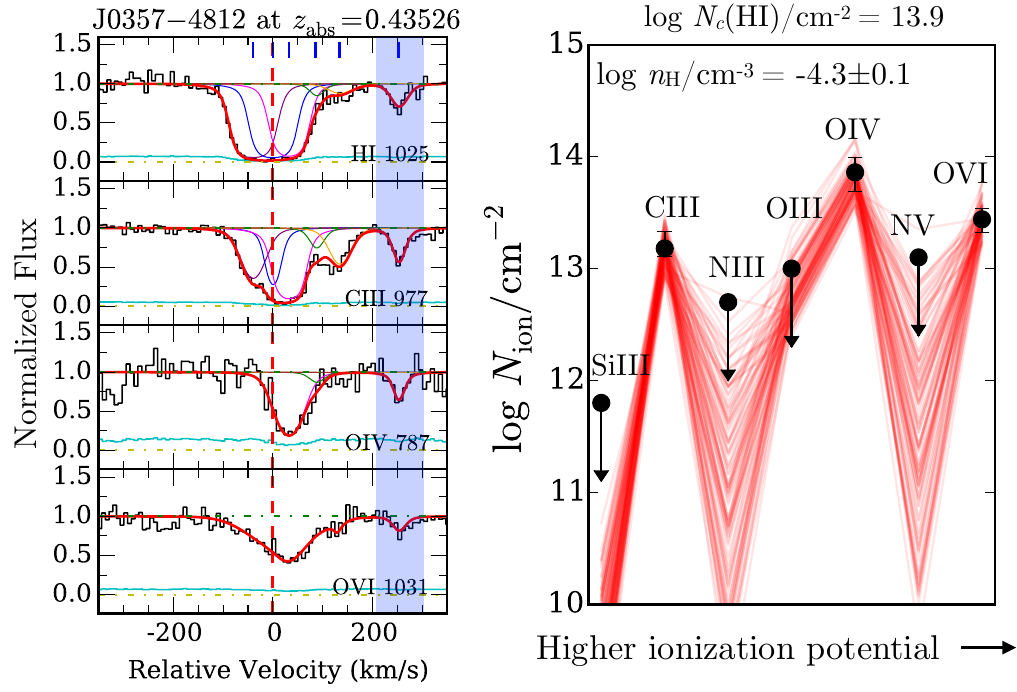}
\vspace{-0.5em}
\caption{{\it Left}: Select absorption transitions for the LLS at $z_\mathrm{abs}=0.43526$ toward J0357$-$4812. The component highlighted in blue exhibits \ion{O}{VI} absorption that is kinematically coincident  with \ion{H}{I} and other low ions. The \ion{O}{VI} linewidth is comparable to those of the lower ions and indicates that the \ion{O}{VI} absorption arises in a cool ($T\sim10^4$\,K), photoionized gas. 
{\it Right}: Comparison between observed column densities (black circles) and model predictions (red curves) for this component. The thin red curves show 100 random MCMC realizations of predicted ionic column densities from \textsc{cloudy} and demonstrate that photoionization models reproduce the observed values well.}

\label{figure:ions}
\end{figure}

A primary source of uncertainty for these inferences is the unknown shape and intensity of the UVB radiation field at $z<1$. 
This uncertainty is driven by the lack of stringent observational constraints on the slope of the UVB at both high (e.g., Becker \etal\ 2015) and low redshifts (e.g., Shull \etal\ 2015). 
Although a number of studies have investigated how uncertainties on the UVB propagate to inferences on gas metallicity and density in the low-redshift CGM/IGM (e.g., Chen \etal\ 2017; Zahedy \etal\ 2019a; Wotta \etal\ 2019), it is also important to explore their impact on the inferred cloud size. Note that these aforementioned studies found that the systematic error on the overall intensity of the UVB is proportional to the uncertainty on the inferred gas density, for a gas that is sufficiently ionized. Because the ionization parameter $U$ is observationally constrained by the ionic column density ratios of different metal species, 
the cloud thickness is then inversely proportional to the overall intensity of the UVB. For example, the inferred median clump thickness for our sample is $\approx250$ pc assuming the FG20 UVB (cf. 160 pc for the HM05 UVB), which is explained by FG20's lower flux of ionizing photons compared to HM05 (see \S 3.2). On the other hand, the possible presence of additional ionizing photons from nearby star-forming galaxies could shift the inferred gas densities to higher values and result in smaller cloud sizes. 

To investigate how uncertainties in the overall intensity of the UVB affect the inferred cloud sizes in our absorber sample, we perform the following Monte Carlo experiment. For each individual component, we allow the overall intensity of the UVB to vary randomly from that of the fiducial HM05 UVB at $z=0.5$ by a factor of up to $\pm0.5$ dex. This choice takes into account not only the differences between UVB models, but also the possible presence of local sources of ionizing radiation (e.g., Schaye 2006; Cantalupo 2010; Rahmati \etal\ 2013; Chen \etal\ 2017; see Appendix A3).\footnote{As noted in Chen \etal\ (2017), a typical $L^*$ star-forming galaxy at $d\lesssim 50$ kpc could provide a comparable or higher number of ionizing photons compared to the UVB.} We repeat this experiment 10000 times, recalculating the PDF of cloud sizes each time. The result of our experiment is summarized by the thin red curves in the right panel of Figure 6, where we plot 100 random realizations of the cumulative PDF of cloud sizes. By allowing the UVB intensity to vary by a factor of $\approx3$ among individual components, we find that the inferred median cloud thickness may vary between $\ell_\mathrm{1/2} \approx100$ pc and $\ell_\mathrm{1/2} \approx 300$ pc. More importantly, this experiment demonstrates that the conclusion that a large majority ($\approx80$ percent) of these cool clumps are spatially compact ($\lesssim$ a few kpc) is robust. 

\section{Discussion}

The main findings of \S\ 4 are as follows. First, individual components associated with the four LLSs comprise relatively quiescent cool gas with $\langle T \rangle =(2\pm1) \times10^4\,$K and modest non-thermal/bulk broadening of $b_\mathrm{nt}\lesssim8\,$\kms. The range of inferred gas densities indicates that these absorbers consist of compact gaseous structures with a median line-of-sight thickness of $\ell_\mathrm{1/2} =160^{+140}_{-50}$ pc. Furthermore, the gas has been chemically enriched to a median metallicity of  $\mathrm{[\alpha/H]_\mathrm{1/2}}=-0.7^{+0.1}_{-0.2}$ and exhibits a wide range of inferred elemental abundance ratios that indicate a diversity of physical origins. Here we incorporate the known galaxy environments of these absorption-line systems and discuss the physical connection between star-forming regions in galaxies and diffuse gas in the gaseous haloes. 

\begin{figure*}
\includegraphics[scale=1.4]{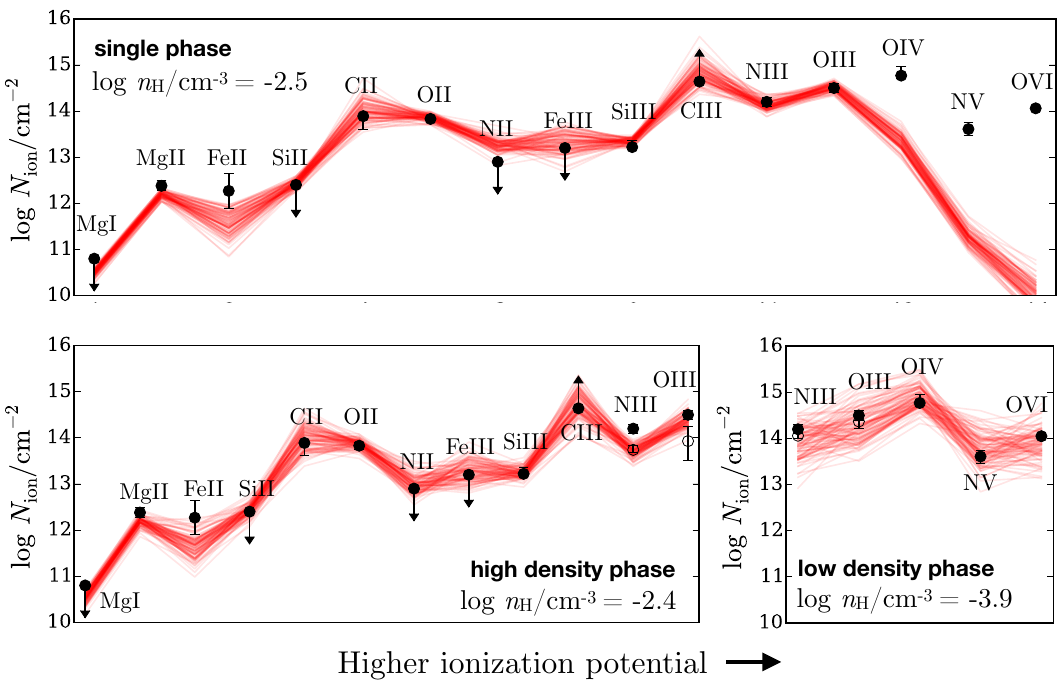}
\vspace{-0.5em}
\caption{
Comparison between observed column densities (black circles) and photoionization model predictions (red curves) from \textsc{cloudy}  for the component at $dv_c=+32$ \kms\ in the LLS toward J0357$-$4812. Symbols have the same meaning as in the right panel of Figure 7. {\it Top}:  A single-phase model is able to reproduce the observed column densities of low-ionization and intermediate-ionization species, but it significantly underpredicts the observed column density of \ion{O}{IV} and other high-ionization species. {\it Bottom}: A two-phase photoionization model comprising a high-density gas phase with $n_\mathrm{H}\approx4\times10^{-3}$ \cmjjj\ and a low-density phase with $n_\mathrm{H}\approx10^{-4}$ \cmjjj\ can reproduce the observed column densities well, including those of high-ionization \ion{N}{V} and  \ion{O}{VI}. Note that for intermediate-ionization species \ion{N}{III} and \ion{O}{III}, the column densities constrained for each phase are shown in open circles, whereas the corresponding total column densities for both phases are shown in black circles. 
}
\label{figure:ions}
\end{figure*}

\subsection {The multiphase nature of the CGM}

All four LLSs in our sample exhibit a multicomponent kinematic structure that is spread over several hundred \kms\ in line-of-sight velocity. Our ionization analysis based on matched component structure in these systems has allowed us to obtain robust constraints on gas densities and chemical abundances in $\approx75$ percent of identified components, with upper/lower limits obtained for the remaining $\approx25$ percent of components. With every LLS,  a single component is found to dominate the total $N\mathrm{(\ion{H}{I})}$ of the system (see discussion in Appendix A). However, these dominant components are not necessarily the most chemically enriched (see also Zahedy \etal\ 2019a). As discussed in \S 4.2, we find no statistically significant correlation between gas metallicity and \ion{H}{I} column density of individual components. On the other hand, components with higher $N_c\mathrm{(\ion{H}{I})}$ tend to also trace denser gaseous structures in the system (Figure 5). At the same time, the wide range of gas densities observed among different components associated with individual LLSs, from $n_\mathrm{H}\approx10^{-4}\, \cmjjj$ to $n_\mathrm{H}\approx0.1\, \cmjjj$, suggests the presence multiphase gas in the CGM. 

The multiphase nature of these LLSs is further indicated by the simultaneous presence of narrow absorption profiles of low ions (like \ion{Mg}{II} and \ion{Fe}{II}) and the broad absorption profiles of more high-ionization species (see also Cooper \etal\ in prep.).  In each of the four LLSs, we detect associated absorption from high-ionization \ion {O}{VI} as well (but see Paper II). Although each of these \ion {O}{VI} absorption profiles exhibits a similar kinematic spread to that of the low-ionization species (see Figures A1 to A4 in Appendix A), which may indicate that they originate within the same halo, the detailed kinematic structure (number of components, velocity centroids, and linewidths) of the \ion{O}{VI} is generally distinct from that of lower ionization species. In particular, the significantly broader line widths of these \ion{O}{VI} absorbers may indicate that the \ion{O}{VI}-bearing gas clouds originate in a distinct gas phase which is more highly ionized, spatially extended, and volume filling than lower ionization gas (see also Zahedy \etal\ 2019a; Rudie \etal\ 2019). 

\begin{figure*}
\includegraphics[scale=1.4]{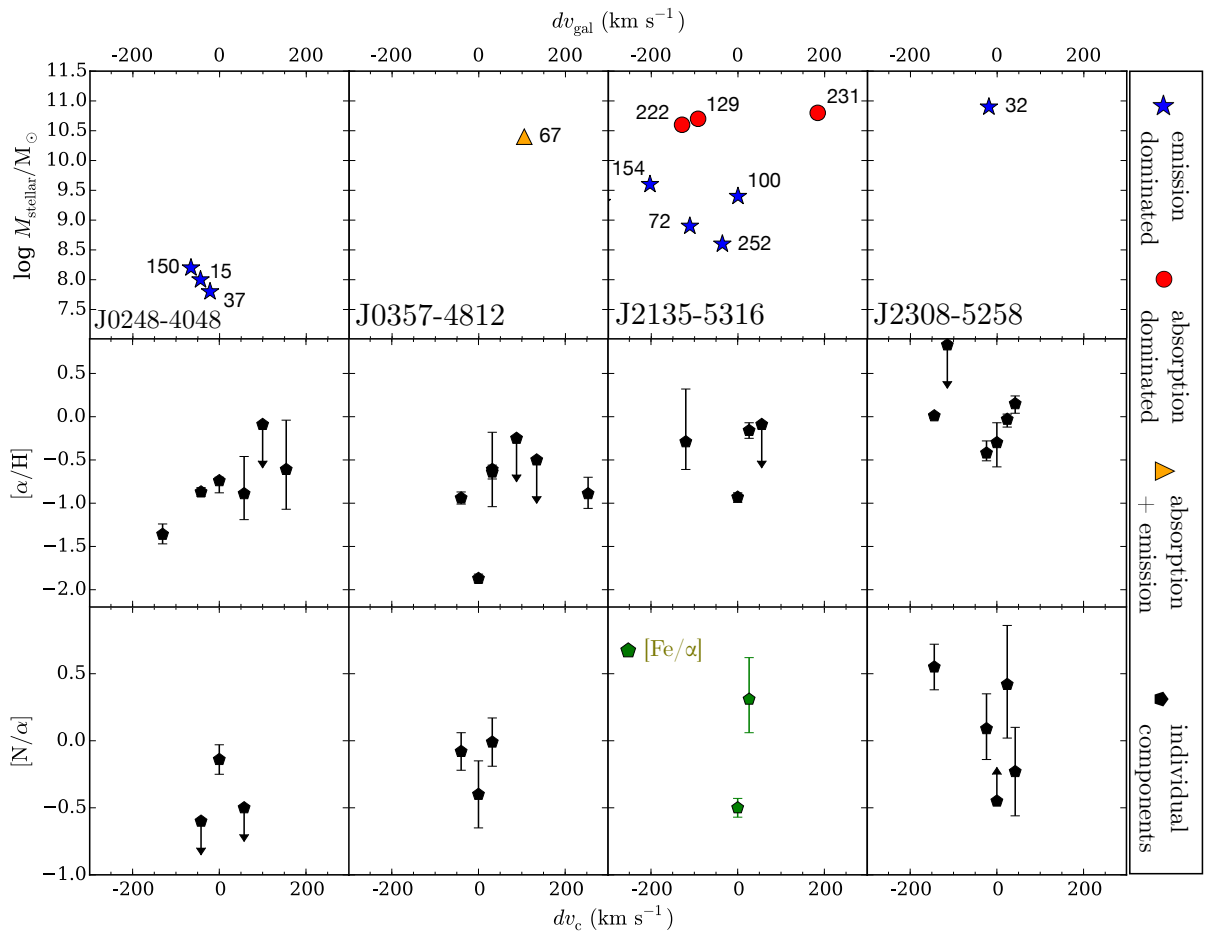}
\vspace{-0.5em}
\caption{
{\it Top}: Galaxy stellar mass versus line-of-sight velocity of spectroscopically identified galaxies  at $d< 300$ kpc and $|d\,v_{\rm gal}|<300$ \kms\ from the CUBS LLSs (see Paper I). Zero velocity corresponds to the line centroid of the strongest (optically thick) \HI\ component of each LLS. 
Emission-line dominated galaxies are denoted by blue stars; absorption-line dominated galaxies in red circles; galaxies displaying both emission and absorption lines in orange triangles. The number adjacent to each data point indicates the projected distance (in kpc) of the galaxy from the absorber. The four LLSs are associated with a diversity of galaxy environments. 
{\it Middle}: Inferred $\mathrm{[\alpha/H]}$ of individual components plotted versus velocity offset $d\,v_c$ from the strongest \HI\ component in each system.
{\it Bottom}: $\mathrm{[N/\alpha]}$ abundance ratio versus $d\,v_c$ for well-constrained components. For  J2135$-$5312, no $\mathrm{[N/\alpha]}$ constraints were possible, so $\mathrm{[Fe/\alpha]}$ is 
shown instead.}
\label{figure:ions}
\end{figure*}

\begin{figure*}
\includegraphics[scale=0.95]{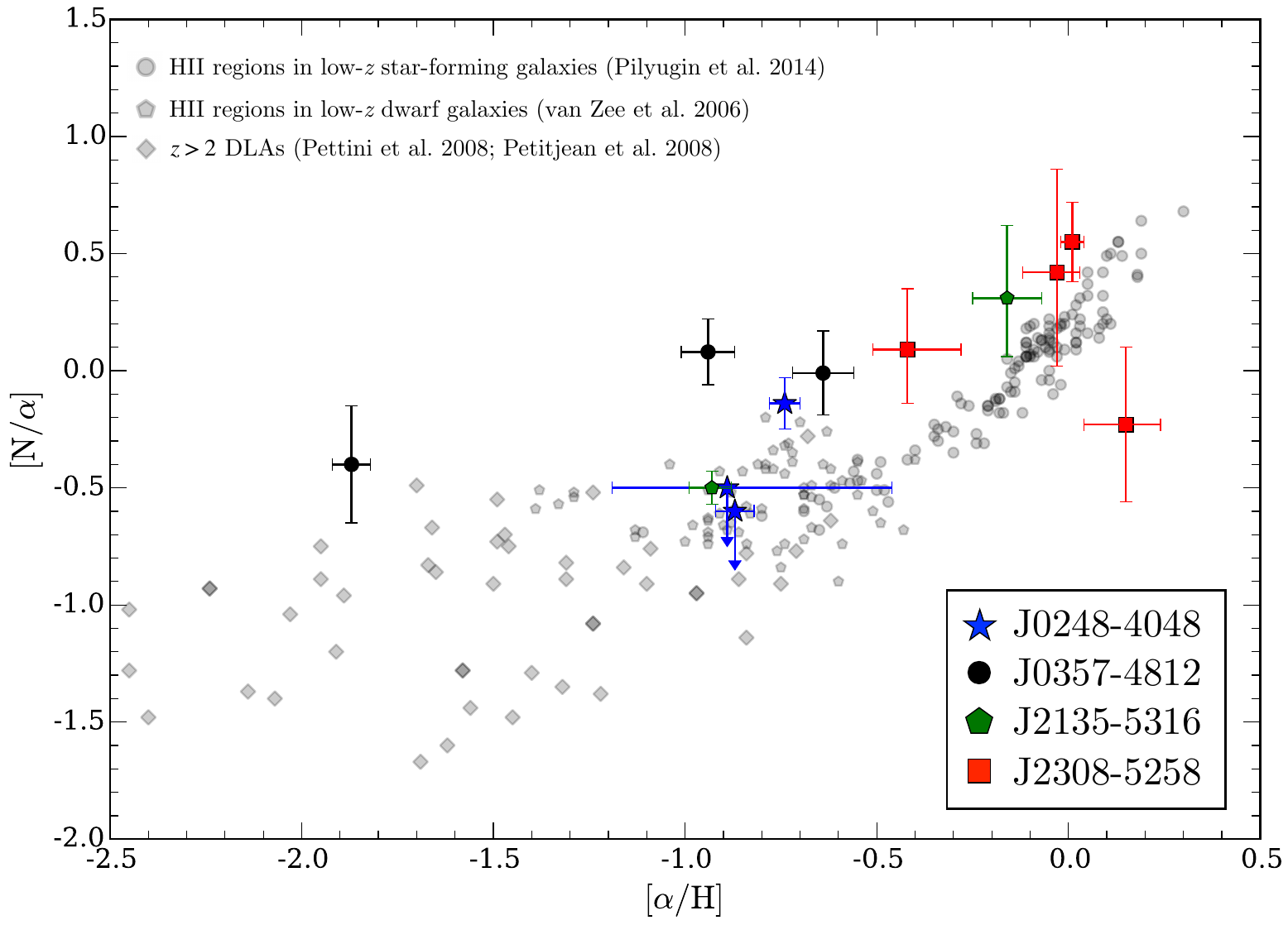}
\vspace{-0.5em}
\caption{
$\mathrm{[N/\alpha]}$ versus gas metallicity for well-constrained components in CUBS LLS sample. No $\mathrm{[N/\alpha]}$ constraints were possible for J2135$-$5312, so $\mathrm{[Fe/\alpha]}$ is 
shown instead. For comparison, we include measurements of gas-phase $\mathrm{[N/\alpha]}$ versus metallicity from individual H\,II regions in both normal star-forming galaxies (Pilyugin \etal\ 2014) and in lower-mass dwarf galaxies (log\,$M_\mathrm{stellar}/\mathrm{M}_\odot\lesssim9$; van Zee \etal\ 2006) in the local Universe, as well as $\mathrm{[N/\alpha]}$  in high-redshift DLAs at $z\gtrsim2$ (Pettini \etal\ 2008; Petitjean \etal\ 2008). Knowledge of both gas metallicity and elemental abundance ratios offers valuable insight into the origin of these LLSs, by drawing a physical connection between star-forming regions in galaxies and diffuse gas in the CGM (see \S 5.2).}
\label{figure:ions}
\end{figure*}

In a minority of cases (four components in two LLSs or $\approx30$ percent of all \ion{O}{VI} components identified), however, we find excellent agreement in the velocity centroids of \ion{O}{VI} and lower ionization species. As illustrated in Figure 7, in each case the \ion{O}{VI} component is relatively narrow ($b_c\lesssim20$ \kms) and has a comparable linewidth to \ion{H}{I}, which indicates that the \ion{O}{VI} absorption originates in a cool  ($T\lesssim$ a few $\times10^4$ K), photoionized gas. A photoionized origin for these narrow \ion{O}{VI} absorbers is further supported by the fact that the observed ionic column densities are well-reproduced by our photoionization modeling (Figure 7). While the sample of photoionized \ion{O}{VI}-bearing gas in our sample is small, our finding suggests that photoionized \ion{O}{VI} is not rare in the $z<1$ CGM, which is in agreement with what has been found empirically at $z<1$ (e.g., Savage \etal\ 2014; Stern \etal\ 2018) and $z\approx2$ (e.g., Carswell \etal\ 2002). 

In most cases, the observed ionic column densities of the observed low-ionization and intermediate-ionization metal species are well reproduced by a single photoionized gas phase. The exception is found with the component at $dv_c=+32$ \kms\ in the LLS at $z_\mathrm{abs}=0.43526$ toward J0357$-$4812. In addition to all the usual low- and intermediate-ionization metals, this component also exhibits kinematically well-aligned \ion{O}{VI} and \ion{N}{V} absorption.  As discussed in Appendix A2 and illustrated in Figure 8, a single-phase model underpredicts the observed \ion{O}{IV}, \ion{N}{V}, and \ion{O}{VI} column densities by more than an order of magnitude. In addition, the observed linewidth of the \ion{O}{IV} component is almost twice as broad as that of the lower-ionization \ion{O}{II} species, a strong indication of the multiphase nature of the gas (Cooper \etal\ in prep). 

In contrast, a two-phase photoionized gas composed of a high-density phase with  $n_\mathrm{H}\approx4\times10^{-3}$ \cmjjj\ and a low-density phase with $n_\mathrm{H}\approx10^{-4}$ \cmjjj\ is able to reproduce all the observed ionic column densities well, including those of the high-ionization \ion{O}{VI} and \ion{N}{V} (Figure 8). However, we note that  the metallicity of the low-density (high-ionization) phase is subject to high uncertainty because the amount of \ion{H}{I} associated with it is not well constrained. However, it is still possible to obtain some constraint on the metallicity  by imposing a physically motivated prior on the cloud size. Specifically, we require the highly ionized phase to be (1) more spatially extended than the denser, lower ionization phase and (2) smaller than $\approx$100 kpc (the size scale of typical gaseous haloes). Based on this prior, the range of allowed gas metallicity for the highly ionized phase is between $\mathrm{[\alpha/H]}\approx-1\,(-1.4)$ and $\mathrm{[\alpha/H]}\approx-0.2\,(0.4)$ at the 68 percent (95 percent) confidence level, which brackets the inferred metallicity of the dense phase, $\mathrm{[\alpha/H]}\approx-0.6$ (Appendix A2). 

The example presented in Figure 8 further underscores the importance of observing ionic absorptions from higher ionization species such as \ion{O}{IV}, \ion{C}{IV}, and  \ion{Si}{IV} (see e.g., Zahedy \etal\ 2019a; Sameer \etal\ 2021) to reveal and characterize multiphase gas  in the CGM. While the gas density and the metallicity of a denser, less ionized gas phase is relatively well constrained by the observed column density ratios of various low ions, additional coverage of these higher ionization species may confirm the presence of more highly ionized gas phases in these absorbers and allow us to better constrain their physical properties. 
 
A particularly noteworthy finding from our study is that the majority of gas clouds associated with the cool phase of LLSs are spatialy compact. As discussed in \S 4.3, 80 percent of these cool clouds have inferred thicknesses of $\lesssim$ a few kpc, with an estimated median size of $\approx 160$ pc. Rauch and collaborators previously investigated spatial variations in the ionic column densities of absorbers detected along multiple sightlines of gravitationally lensed QSOs, and concluded that cool gas clumps traced by \ion{Mg}{II} and \ion{Fe}{II} species have size upper limits of $\lesssim100-200$ pc  (e.g., Rauch \etal\ 1999; 2002; see also Ellison \etal\ 2004). Our estimates, which rely on photoionization modeling, are in excellent agreement with the model-independent constraints of transverse cool cloud size obtained using gravitational lenses. The small sizes of these cool clouds, along with the possible evidence that more highly ionized gas arises in more spatially extended structures (Figure 5), suggest that the CGM consists of a large number of compact, cool clouds which are embedded in highly ionized and extended structures that are likely also more volume filling (see also Rauch \etal\ 2001; Schaye \etal\ 2007; Rudie \etal\ 2019).

\subsection {On the physical origins of LLS absorbers}

Zahedy \etal\ (2019a) previously reported that cool ($T\sim10^4$ K) gas metallicity and number density can vary by more than an order of magnitude within individual haloes of massive elliptical galaxies with log\,$M_\mathrm{stellar}/\mathrm{M}_\odot>11$. These authors interpreted their findings as evidence that the CGM of massive galaxies comprises a multiphase mixture of gases with different chemical enrichment histories. Similar to the findings of Zahedy \etal\ (2019a), our detailed ionization analysis has uncovered large variations in gas density and metallicity among different components associated with individual LLSs in the CUBS sample (\S 4.2; Figure 9, middle). The consistent findings between these two studies indicate that large variations in chemical abundances and densities are general features of the CGM instead of specific characteristics of massive ($M_\mathrm{h}\sim10^{13}\,\mathrm{M}_\odot$) dark-matter haloes hosting luminous red galaxies. Furthermore, the observed large fluctuations in gas metallicity within individual haloes suggest that chemical mixing in the CGM is an inefficient process and hint at a diverse chemical enrichment histories of the gas. They also underscore the fact that gas metallicity alone provides an incomplete diagnostic of the physical origin of chemically enriched gas in the CGM. From a practical standpoint, these findings highlight the importance of using high-resolution spectra to resolve the kinematic structure of CGM absorbers and recover information on intra-halo variations in elemental abundances.

At the same time, our analysis has also revealed a wide range of elemental abundance ratios within individual absorption systems and among different LLSs studied (Figure 9, bottom), which further supports the interpretation that different physical processes contribute to the chemical enrichment history of the gas. In Paper I, we reported that these LLSs are associated with a wide variety of galaxy environments, based on deep galaxy survey data of these fields obtained with VLT-MUSE and the Magellan Telescopes (see Figure 9). Here we discuss the physical connection between the galaxy environment of these LLSs and the observed physical properties and detailed elemental abundances of the gas. 

\subsubsection {LLS at $z_\mathrm{abs}=0.36400$ toward J0248$-$4048}

This LLS is associated with a pair of low-luminosity ($\approx0.01-0.03\,L^*$) and low-mass (log\,$M_\mathrm{stellar}/\mathrm{M}_\odot<8.2)$ galaxies at projected distances $d=15$ and 37 kpc, respectively (Figure 9). In addition to this pair, a third dwarf galaxy is situated farther away at $d=150$ kpc from the LLS. The gas-phase metallicities of these galaxies are estimated to be $\approx 0.1$ solar (Paper I). 

We obtain robust constraints on the gas metallicity for five out of six individual components in this LLS. While there is a spread in $\mathrm{[\alpha/H]}$ among different components (Appendix A1), the gas has a consistently sub-solar metallicity with $\mathrm{[\alpha/H]}<-0.5$, similar to the estimated gas-phase metallicities in the ISM of the two nearest galaxies. Furthermore, we find  sub-solar values of $\mathrm{[N/\alpha]}\lesssim-0.2$ in all components  where robust constraints can be obtained. As shown in Figure 10, the range of inferred gas metallicities and $\mathrm{[N/\alpha]}$ ratios in this LLS are similar to the observed $\mathrm{[N/\alpha]}$ ratios in individual \ion{H}{II} regions of low-mass galaxies in the nearby Universe (van Zee \etal\ 2006). 

\begin{figure}
\includegraphics[scale=0.63]{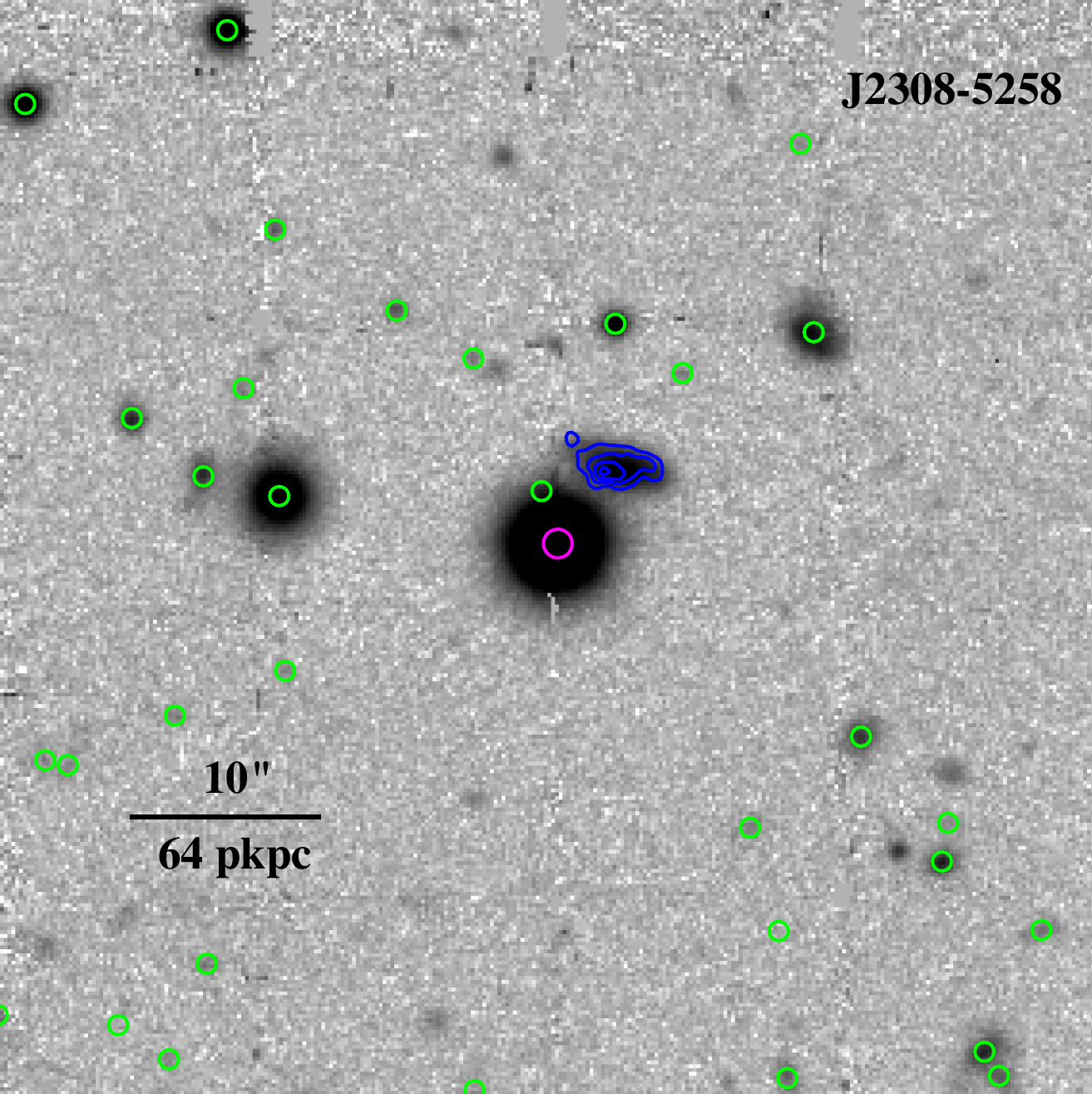}
\caption{
Galaxy environment associated with the LLS at $z_\mathrm{abs}=0.54273$ toward J2308$-$5258. The pseudo $r$-band image is produced by integrating the MUSE cube over the wavelength range $6000-7000$ A. North direction is up, while East is to the left. Out of the 34 objects spectroscopically identified within this image (green circles), only one galaxy (blue contour) is situated near the LLS, at $d=32$ kpc and  $dv_\mathrm{gal}=-19$ \kms. The $\mathrm{H\beta}$ emission contours of the galaxy are superimposed on this image, with each curve corresponding to a constant surface brightness level of 0.15, 0.45, 0.75, and 1.05 $\times 10^{-17}\,{\rm erg}\,{\rm s}^{-1}\,{\rm cm}^{-2}\,{\rm arcsec}^{-2}$, respectively. The QSO sightline intercepts the gaseous halo of this galaxy at an orientation angle $\phi\approx 62^\circ$, which is $<30^\circ$ from the galaxy's minor axis.
}
\label{figure:ions}
\end{figure}

The similarities between $\mathrm{[N/\alpha]}$ ratios and metallicities in this LLS and the expected chemical enrichment levels in the ISM of low-mass galaxies indicate a physical connection between the LLS and the two proximate galaxies. At the same time, we further note that two components exhibit high values of $\mathrm{[Fe/\alpha]}$ ratios, $\mathrm{[Fe/\alpha]}\approx0.2-0.3$ (Figure 4, left). High $\mathrm{[Fe/\alpha]}$ ratios are expected in galaxies which have been continuously forming stars over time, allowing their gas to have been substantially enriched by SNe Ia. The fact that they occur at low absolute metallicities ($\approx0.1-0.2$ solar) suggests that the chemical buildup process in these dwarf galaxies is relatively slow and inefficient, which points to either overall low time-averaged star formation rates in these galaxies or significant metal loss in SNe CC-driven galactic winds throughout their history. We conclude that the LLS originates from gas expelled from the ISM of the nearby pair of dwarf galaxies, in particular tidally stripped gas during past or ongoing interactions between the two galaxies (e.g., Stierwalt \etal\ 2015; Pearson \etal\ 2016). While it is also possible that the gas comes from a starburst-driven wind out of one or both galaxies, we consider such a scenario less likely in light of the inferred high $\mathrm{[Fe/\alpha]}$ relative abundances in these two components (e.g., Konami \etal\ 2011; Mitsuishi \etal\ 2013). 

\subsubsection {LLS at $z_\mathrm{abs}=0.43526$ toward J0357$-$4812}

In contrast to the pair of dwarf galaxies found around the LLS toward J0248$-$4048, the LLS at $z_\mathrm{abs}=0.43526$ toward J0357$-$4812 is situated in a relatively isolated galaxy environment. Specifically, our deep galaxy survey has identified only one luminous galaxy within $d<300$ kpc (Figure 9). This galaxy is a $\approx1.4\,L^*$ star-forming galaxy (log\,$M_\mathrm{stellar}/\mathrm{M}_\odot=10.4)$ at $d=67$ kpc and  $dv_\mathrm{gal}=+67$ \kms\ from the LLS (Paper I). The QSO sightline intercepts the gaseous halo of this galaxy at an orientation angle $\phi\approx27^\circ$ from the galaxy's major axis. 

The observed line-of-sight kinematics of the three strongest components of this LLS, which contain $>99$ percent of the total $N\mathrm{(\ion{H}{I})}$, are consistent with the expectation for a gas that co-rotates with the galaxy disk (see Figure 10 of Paper I). Our ionization analysis indicates that these three dominant components likely arise in a dynamically cold gaseous structure, based on their observed narrow linewidths and low turbulent line broadening of $b_\mathrm{nt}\lesssim4$\,\kms. Furthermore, all three components have sub-solar metallicities of $\mathrm{[\alpha/H]}<-0.7$, with the optically thick component being the most metal-poor with a metallicity of $\approx0.01$ solar. The metal deficient nature of the optically thick component is confirmed by the low $\mathrm{[N/\alpha]}$ ratio, $\mathrm{[N/\alpha]}\approx-0.40\pm0.25$, which underscores the chemically primitive nature of the gas. The observed  $\mathrm{[N/\alpha]}$ abundance ratio is comparable to those observed in high-redshift DLAs (e.g., Pettini \etal\ 2008; Petitjean \etal\ 2008), and suggests that the gas has been mostly enriched by primary nitrogen produced in earlier generations of stars (e.g., Spite \etal\ 2005). This early-enrichment scenario is consistent with what is expected from recently accreted, chemically primitive gas from the IGM (e.g., Rauch \etal\ 1997).

On the other hand, the other two strong components exhibit  $\mathrm{[N/\alpha]}$ values of $\mathrm{[N/\alpha]}\approx0$, which are in contrast to their inferred sub-solar metallicities of $0.1-0.2$ solar. This is rather surprising because the inferred $\mathrm{[N/\alpha]}$ ratios are more similar to what is expected in the ISM of the $L^*$ disk galaxy associated with the LLS and indicates a chemically mature system (Figure 10), as corroborated by the similarly high inferred $\mathrm{[C/\alpha]}$. How can we reconcile these high  $\mathrm{[N/\alpha]}$ and $\mathrm{[C/\alpha]}$ abundance ratios with the otherwise low metallicity of the gas? One possible explanation for the origin of these two components is that the gas was ejected from the disk galaxy at a high, near-solar metallicity (thereby explaining the high $\mathrm{[N/\alpha]}$ ratios), but it has since been mixed with and diluted by more metal-poor and chemically primitive material in the CGM. Such dilution by metal-poor gas could have reduced the overall gas metallicity without significantly changing the overall elemental abundance ratios of the gas (e.g., Frye \etal\ 2019). In conclusion, we consider a combination of recycled gas and more chemically primitive gas accreted from the IGM to be the most likely explanation for the observed LLS toward J0357$-$4812. 

\subsubsection {LLS at  $z_\mathrm{abs}=0.62255$ toward J2135$-$5316}

The LLS at  $z_\mathrm{abs}=0.62255$ toward J2135$-$5316 is associated with a massive galaxy group with an estimated dynamical mass of $M_{\rm dyn}\approx 1.1\times 10^{13}\,\msun$ (Figure 9; Paper I). We obtain robust constraints on gas metallicity and the $\mathrm{[Fe/\alpha]}$ and $\mathrm{[C/\alpha]}$ abundance ratios for two (out of six) individual components of this LLS. Specifically, the optically thick component has a low metallicity of $\mathrm{[\alpha/H]}\approx-0.9$ and exhibits an $\alpha$-rich elemental abundance pattern of $\mathrm{[Fe/\alpha]}\approx-0.5$. The low value of $\mathrm{[Fe/\alpha]}$ indicates a nucleosynthetic origin that is dominated by core-collapse SNe (e.g., Nomoto \etal\ 2006) and is comparable to the typical stellar $\mathrm{[Fe/\alpha]}$ in massive quiescent galaxies. However, the predominantly old stellar populations in these massive systems also exhibit near-solar (or higher) metallicities (e.g., Greene \etal\ 2015). When considered with the low overall gas metallicity, a possible explanation is that the absorbing gas originates from the stripped ISM or outflowing gas out of a lower-mass members of the galaxy group. The closest galaxy in projection at $d=72$ kpc is one such low-mass galaxy (log\,$M_\mathrm{stellar}/\mathrm{M}_\odot\approx9$; Figure 9), with an inferred gas-phase ISM metallicity (following the mass-metallicity relation of Berg \etal\ 2012) that is comparable to that of the optically thick component. 

In contrast, the other component with robust chemical abundance constraints in this LLS is characterized by both a higher gas metallicity, $\mathrm{[\alpha/H]}\approx-0.2\pm0.1$, and elemental abundance ratios, $\mathrm{[Fe/\alpha]}\approx0.2\pm0.2$ and  $\mathrm{[C/\alpha]}\approx0.3\pm0.2$. These values indicate that this component originates in a chemically evolved gas which has been subjected to substantial enrichment from later generations of stars (e.g., McWilliam 1997; Gustafsson \etal\ 1999). Given that our deep galaxy survey of this field had previously identified 5 massive galaxies with log\,$M_\mathrm{stellar}/\mathrm{M}_\odot=10.2-11.2$ at $d\lesssim250$ kpc from the LLS, it is likely that the gas associated with this absorption component was originally located in the ISM of one of these galaxies but has since been expelled into its CGM or the intra-group medium of this massive group. 

\subsubsection {LLS at $z_\mathrm{abs}=0.54273$ toward J2308$-$5258}

Similar to the LLS toward J0357$-$4812, the LLS at $z_\mathrm{abs}=0.54273$ toward J2308$-$5258 is situated in a relatively isolated galaxy environment. 
The LLS is associated with a luminous ($\approx 2\,L^*$) and massive (log\,$M_\mathrm{stellar}/\mathrm{M}_\odot=10.9$) star-forming galaxy situated at $d=32$ kpc and $dv_\mathrm{gal}=-19$ \kms\ (Figure 9).  The QSO sightline intercepts the gaseous halo of this inclined galaxy at $\phi\approx 62^\circ$ from the major axis, as shown by the deep pseudo $r$-band image from MUSE presented in Figure 11. Given the background sightline's proximity to the projected minor axis of the galaxy, which has been claimed to exhibit an elevated incidence of outflowing \ion{Mg}{II}-bearing gas in the CGM (e.g., Bouch{\'e} \etal\ 2012), this LLS provides a valuable opportunity to test the assumptions of this simple picture of a biconical outflow with the observed elemental abundances. 

We are able to constrain the elemental abundances in five out of six individual components in this LLS. While there is a spread of 0.6 dex in the inferred $\mathrm{[\alpha/H]}$ among components (see Appendix A4), the gas exhibits a high degree of enrichment, with metallicities of $\approx0.4-1.2$ solar. In four of these components, the high metallicities are accompanied by inferred elemental abundance ratios of $\mathrm{[Fe/\alpha]}\approx0-0.5$, $\mathrm{[C/\alpha]}\gtrsim0.2$, and $\mathrm{[N/\alpha]}\gtrsim0-0.4$. The high gas metallicities and elemental abundance ratios indicate a chemically evolved system, which is consistent with a gas that originates from the ISM of the luminous star-forming galaxy (see Figure 10). 

In contrast, the fifth component paints a different story. This component, which has a log\,$N\mathrm{(\ion{H}{I})/\cmjj}=16.5\pm0.1$, is one of the most metal-rich components of the LLS, with $\mathrm{[\alpha/H]}=0.0\pm0.1$. At the same time, the component exhibits an Fe-deficient chemical enrichment signature with  
$\mathrm{[Fe/\alpha]}\approx-0.4\pm0.1$. The inferred $\mathrm{[Fe/\alpha]}$ for this component suggests a chemical enrichment history that is dominated by core-collapse SNe, and is consistent with observations of outflowing gas around local starburst galaxies (e.g., Konami \etal\ 2011; Mitsuishi \etal\ 2013), where $\mathrm{[Fe/\alpha]}$ ratios between $-0.5$ and $-0.3$ are common. It is therefore likely that this component originates in outflowing gas from the nearby massive star-forming galaxy. However, the fact that the other four components show more chemically evolved enrichment signatures suggest that this possible galactic wind is not only expelling metals recently produced in SNe CC, but also other preexisting materials from the ISM of the galaxy.  

In summary, we have considered the physical connection between the galaxy environment of LLSs and the observed physical properties and elemental abundances of the gas. While the sample size of absorption systems for which this investigation has been done is small (four systems), we have demonstrated that coupling the information about gas metallicity with the known elemental abundance ratios of the gas provides a powerful tool to resolve the origin of chemically enriched gas in CGM.

\section{Summary and Conclusions}

We have carried out a systematic investigation of the physical conditions and chemical enrichment in four new LLSs at $z=0.36-0.6$ discovered within the Cosmic Ultraviolet Baryonic Survey (CUBS). CUBS consists of 15 QSOs selected based on their NUV brightness, enabling an unbiased search of intervening LLSs at $z<1$. The high quality of the QSO absorption spectra has allowed us to to infer the physical conditions of the gas, using a detailed ionization modeling that takes into account the resolved component structures of \ion{H}{I} and various low- and high-ionization metal transitions. Our main findings are summarized below. 

1) All four LLSs exhibit multi-component kinematic structures, with
associated metal transitions from multiple ionization states such as
\HI, \CII, \CIII, \MgII, \FeII, and \OVI\ absorption that span $\approx200-350$ \kms\ in line-of-sight velocity. The multiphase nature of these LLSs is indicated by the simultaneous presence of narrow absorption lines of low-ionization metal species and the broad absorption profiles of more highly ionized species. Specifically, each LLS has associated absorption from highly ionized \ion {O}{VI}, which exhibits a kinematic structure (number of components, their velocity centroids and linewidths) that is generally distinct from that of lower ionization species. 

2) \ion{H}{I}-bearing gas associated with LLSs is cool, with $T<10^5$ K inferred from the observed \ion{H}{I} linewidths. The observed linewidth ratios for matched \ion{Mg}{II} and \ion{H}{I} components indicate that gas clouds associated with these higher column density components (log\,$N$(\ion{H}{I})/\cmjj$ \gtrsim16$) have a mean temperature and dispersion of $\langle T \rangle =(2\pm1) \times10^4\,$K and a non-thermal line broadening of  $\langle b_\mathrm{nt} \rangle =5\pm3\,$\kms (Figure 2). 

3) By considering matched kinematic components and comparing the relative abundances of different ions to predictions from photoionization models, we find that both metallicity and gas number density can vary by a factor of five or more among different components within individual LLSs (Figure 3).  These large fluctuations suggest that chemical mixing in the CGM is an inefficient process and also point to a diversity of chemical enrichment histories of the gas. Furthermore, they underscore that gas metallicity alone is an incomplete diagnostic of the physical origins of chemically enriched gas in the CGM. Pragmatically, they highlight the importance of using high-resolution spectra to resolve the kinematic structure of CGM absorbers.  

4) While obtaining robust metallicity constraints for the low-density, highly ionized gas phase is challenging due to the uncertain $N$(\ion{H}{I}), we estimate that the median metallicity of absorbing components in the denser, lower-ionization phase is $\mathrm{[\alpha/H]_\mathrm{1/2}}=-0.7^{+0.1}_{-0.2}$, with a 16-84 percentile range of $\mathrm{[\alpha/H]}=(-1.3,-0.1)$. Two out of four LLSs also exhibit metal-poor components with $\mathrm{[\alpha/H]}<-1$. 

5) We find a wide range of inferred relative elemental abundance ratios ($\mathrm{[C/\alpha]}$, $\mathrm{[N/\alpha]}$, and $\mathrm{[Fe/\alpha]}$) within and among different LLSs, which indicates a diversity of chemical enrichment histories (Figure 4 and Figure 10). 

6) The median gas density is log\,$n_\mathrm{H}/ \cmjjj\approx-2.4$ for individual components assuming the HM05 UVB (or log\,$ n_\mathrm{H}/ \cmjjj\approx-2.6$ under FG20), which translates to a median ionization parameter of $\mathrm{log}\, U_\mathrm{1/2}\approx-2.9$ . The data also exhibit a trend of rising gas density (declining ionization parameter) with increasing $N$(\ion{H}{I}) (Figure 5). The median ionization parameter for absorption components with log\,$N_c\mathrm{(\ion{H}{I})/\cmjj}\gtrsim16$ is lower,  $\mathrm{log}\,U_\mathrm{1/2}\approx-3.3$, in agreement with previous surveys of partial/full LLSs at $z<1$. 

7) Cool clouds associated with $z<1$ LLSs are spatially compact. The inferred cloud sizes are between 10 pc and $\approx10$ kpc with a mode of $\sim100$ pc (Figure 5), and $\approx80$ percent of these cool clouds have an inferred thickness of $\lesssim$ a few kpc. The estimated median clump size is $\ell_\mathrm{1/2} =160^{+140}_{-50}$ pc (Figure 6), where the quoted uncertainties reflect possible variations in the ionizing background radiation field at $z<1$ (Figure 6).  

8) Combining the absorption data with deep galaxy survey data characterizing the galaxy environment of the four LLSs, we explore the physical connection between star-forming regions in galaxies and the observed physical properties and elemental abundances of the gas (both absolute metallicities {\it and} relative abundance ratios; \S 5.2 and Figure 10). The four LLSs are associated with a diverse range of galaxy environments, ranging from the gaseous haloes of isolated $L^*$ galaxies, to the vicinity of a pair of low-mass galaxies, to a massive galaxy group (Figure 9). By examining the gas metallicities and elemental abundance ratios of individual absorption components associated with the four LLSs, we conclude that the gas originates from a variety of physical processes. These diverse physical origins include gas tidally stripped from the ISM of a pair of interacting dwarf galaxies, co-rotating halo gas which has been mixed with and diluted by metal-poor and chemically primitive material in the CGM, and star-formation driven wind  from an $L^*$ galaxy that also expels chemically mature, preexisting gas in the galaxy's ISM. The wide range of physical origins inferred for LLSs in the CUBS sample underscores the importance of utilizing deep galaxy survey data to fully resolve the nature of chemically enriched gas in the CGM.

In conclusion, we continue to build a rich dataset with CUBS in order to probe the co-evolution of the CGM and galaxies at $z\apl 1$. With the current work, we have demonstrated that combining deep galaxy survey data with knowledge of gas metallicities and elemental abundance ratios provides a powerful tool to resolve the physical origin and chemical enrichment history of CGM absorbers. In future papers from our collaboration (Cooper \etal\ in prep; Johnson \etal\ in prep), we will continue to investigate the physical conditions, chemical compositions, and kinematics of the gaseous haloes around galaxies with diverse star formation histories, examine the physical origins of absorption-line systems in the CGM, and explore their co-evolution with galaxies across cosmic time.

\section*{Acknowledgments}

The authors thank an anonymous referee for 
reviewing our paper and helping us improve the quality
of its presentation.
FSZ is grateful for the support of a Carnegie Fellowship from the
Observatories of the Carnegie Institution for Science. 
HWC, EB, and MCC acknowledge partial support
from HST-GO-15163.001A and NSF AST-1715692 grants.  TC and GCR
acknowledge support from HST-GO-15163.015A.  
KLC acknowledges partial support from NSF AST-1615296. CAFG was supported by NSF through grants AST-1715216 and 
CAREER award AST-1652522; by NASA through grant 17-ATP17-0067; by STScI 
through grant HST-AR-16124.001-A; and by the Research Corporation for Science Advancement through a 
Cottrell Scholar Award and a Scialog Award. SL was funded by project FONDECYT 1191232. 
This work is based on observations
made with ESO Telescopes at the Paranal Observatory under programme ID
0104.A-0147(A), observations made with the 6.5m Magellan Telescopes
located at Las Campanas Observatory, and spectroscopic data gathered
under the HST-GO-15163.01A program using the NASA/ESA Hubble Space
Telescope operated by the Space Telescope Science Institute and the
Association of Universities for Research in Astronomy, Inc., under
NASA contract NAS 5-26555.  This research has made use of NASA's
Astrophysics Data System and the NASA/IPAC Extragalactic Database
(NED) which is operated by the Jet Propulsion Laboratory, California
Institute of Technology, under contract with the National Aeronautics
and Space Administration.

\section*{Data Availability}

The data relevant to this article are presented in Appendix A. Additional data not presented here will be shared on reasonable request to the corresponding author.

\appendix

\section{Notes on Individual systems}

\subsection{LLS at $z_\mathrm{abs}=0.36400$ toward J0248$-$4048}

This LLS is situated at $z_\mathrm{abs}=0.36400$. As shown in Figure A1, the absorber has a total $N\mathrm{(\ion{H}{I})}$ of log\,$N\mathrm{(\ion{H}{I})/\cmjj}=17.57\pm0.01$. The HI absorption is accompanied by secure ionic metal detections of \ion{C}{II}, \ion{C}{III},  \ion{N}{II},  \ion{N}{III}, \ion{O}{I}, \ion{Mg}{I}, \ion{Mg}{II}, \ion{Si}{II}, \ion{Si}{III}, \ion{Fe}{II}, \ion{Fe}{III},  \ion{S}{III}, and \ion{O}{VI}.

Our Voigt profile analysis identifies a minimum of six components in the absorption system, as seen clearly in the absorption panels shown in Figure A1. The absorber has a velocity spread of $\Delta v\approx290$ \kms\ from its bluest to reddest component. The strongest \ion{H}{I} absorption occurs in component 3 at $z_\mathrm{abs}=0.36400$. This component comprises nearly 90 percent of the $N\mathrm{(\ion{H}{I})}$ of the absorber, with most of the remaining \ion{H}{I} column density situated in component 2 at $dv_c=-42$ \kms. 
At the same time, the higher resolution $({\rm FWHM}\approx7\, \kms)$ MIKE spectra of the \ion{Mg}{II}, \ion{Fe}{II}, and \ion{Mg}{I} species reveal that the strongest component at $dv_c\approx0$ \kms\ is actually separable into two distinct kinematic components of comparable absorption strengths that are separated by $<3$ \kms\ in velocity space. Because of this small velocity separation, the limited resolution of our COS spectrum $({\rm FWHM}\approx20\, \kms)$, and the fact that all available \ion{H}{I} transitions are saturated at this location, we cannot reliably constrain the individual \ion{H}{I} or metal column densities of these components from the COS data. However, the sum of  \ion{H}{I} column densities of these two components is still well constrained by the observed flux decrement at the Lyman limit. For that reason, we proceed to perform the photoionization modeling these two low-ionization components together as a single component using the sum of their column densities. This choice is justified by the nearly identical observed $N_c\mathrm{(\ion{Fe}{II})}/N_c\mathrm{(\ion{Mg}{II})}$ of the two components, which indicates that they likely have similar chemical enrichment histories and ionization states. 
 
As shown in Figure A2, the component structure of \ion{H}{I} is in good agreement with that of the metal ions (e.g., \ion{Mg}{II} and \ion{Fe}{II}). The observed \ion{H}{I} linewidths for all components constrain the gas temperature to $T\lesssim 10^5$ K. For components also detected in \ion{Mg}{II}, the ratio of the Doppler linewidths of \ion{H}{I} and \ion{Mg}{II} in each component constrains the gas temperature to $T= (2-3)\times10^4$ K and indicates a modest non-thermal line broadening of $b_\mathrm{nt}\approx5-8$\,\kms.

The well-matched component structure of \ion{H}{I} and low- to intermediate-ionization metals justifies modeling them with a single-phase photoionization model . Our analysis finds that the observed ionic column densities can be reproduced by a photoionized gas spanning a wide range of gas densities among the different components, from log\,$n_\mathrm{H}/ \cmjjj\approx-3$ to log\,$n_\mathrm{H}/ \cmjjj\approx-1.7$ . In contrast, our analysis results indicate a smaller variation in chemical abundance, which ranges from between $\mathrm{[\alpha/H]}=-0.6\pm0.4$ (component 6) and $\mathrm{[\alpha/H]}=-1.3\pm0.1$ (component 1). The modest level of chemical enrichment in the gas is consistent with the inferred values of $\mathrm{[C/\alpha]}$ and  $\mathrm{[N/\alpha]}$ abundance ratios, which are primarily subsolar.  

This LLS also exhibits strong absorption of high-ionization \ion{O}{VI}. The \ion{O}{VI} absorption consists of a minimum of four broad components ($b_c\gtrsim30$ \kms) with a total log\,$N\mathrm{(\ion{O}{VI})/\cmjj}=14.8\pm0.1$.

\begin{figure*}
\includegraphics[scale=1.4]{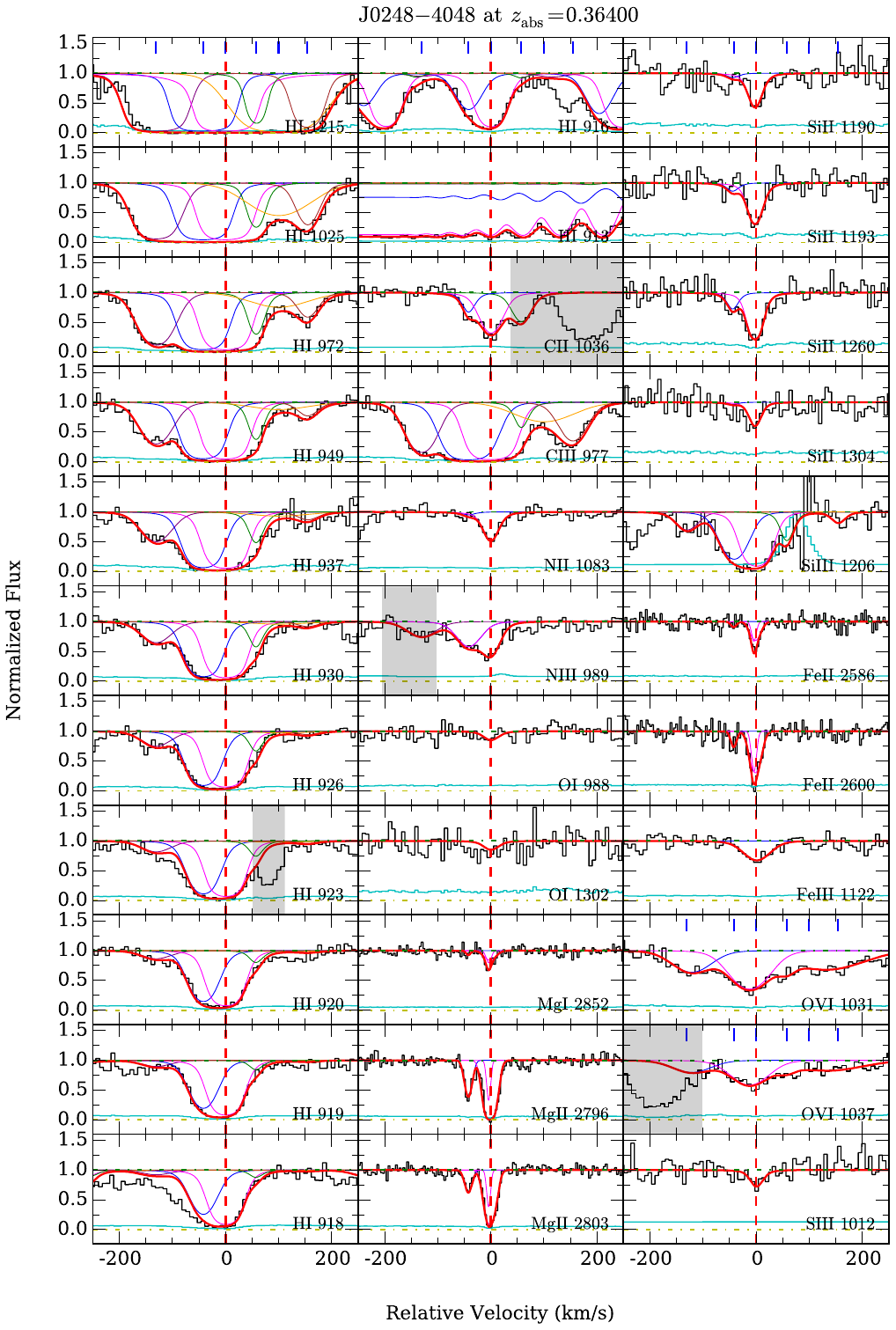}
\vspace{-0.75em}
\caption{Continuum normalized absorption profiles of different transitions of the LLS at $z_\mathrm{abs}=0.36400$ toward J0248$-$4048. Zero velocity indicates the line centroid of the strongest \ion{H}{I} absorption component.
For each transition, the 1-$\sigma$ error spectrum is shown in a cyan-colored curve.
Contaminating features are grayed out, for clarity. The best-fit Voigt profiles are plotted for each detected transition, 
both for individual components (different colored curves) and the sum of all components (red curve). Blue tick marks on the top panels indicate the centroid of each kinematic component. Our higher resolution MIKE data show that the absorption at $dv_c\approx0$ \kms\ is separable into two distinct components as seen in the \ion{Mg}{II}, \ion{Fe}{II}, and \ion{Mg}{I} profiles, with these components shown in the two magenta colored curves (see text in Appendix A1).}
\label{figure:ions}
\end{figure*}

\subsection{LLS at $z_\mathrm{abs}=0.43526$ toward J0357$-$4812}

This LLS is situated at $z_\mathrm{abs}=0.43526$ and has a total $N\mathrm{(\ion{H}{I})}$ of log\,$N\mathrm{(\ion{H}{I})/\cmjj}=17.18\pm0.01$. The following metal species are also present: \ion{C}{II}, \ion{C}{III}, \ion{N}{III}, \ion{N}{V},  \ion{O}{II},  \ion{O}{III},  \ion{O}{IV}, \ion{O}{VI}, \ion{Mg}{II}, \ion{Si}{III}, \ion{Fe}{II}, and \ion{Fe}{III} (see Figure A2). 

Our Voigt profile analysis identifies a minimum of six components in the absorption system, as seen clearly in Figure A2.. The observed full kinematic spread of the absorber is $\Delta v\approx290$ \kms. The strongest \ion{H}{I} absorption occurs in component 2 at $z_\mathrm{abs}=0.43526$, which comprises nearly 80 percent of the system's total $N\mathrm{(\ion{H}{I})}$. Most of the remaining H\,I column density is distributed equally between two components immediately adjacent to it. As shown in Figure A2, the \ion{H}{I} component structure agrees very well with that of the low and intermediate metal ions (e.g., \ion{Mg}{II} and \ion{Fe}{II}). The observed \ion{H}{I} linewidths for all components constrain the gas temperature to $T\lesssim5\times 10^4$ K. For components also detected in \ion{Mg}{II}, the ratio of the Doppler linewidths of \ion{H}{I} and \ion{Mg}{II} in each component constrains the gas temperature to $T= (2-3)\times10^4$ K, with relatively little non-thermal line broadening of  $b_\mathrm{nt}\lesssim4$\,\kms.

This LLS is unique in our sample because the observed ionic column densities in one of its components, component 3 at $dv_c=+32$ \kms, cannot be well-reproduced by a single-phase photoionized gas. 
As shown in Figure 7, a single phase model underpredicts the observed \ion{O}{IV} column density by more than an order of magnitude. To reproduce the observed ionic column densities requires a two-phase gas with  $n_\mathrm{H}\approx4\times10^{-3}$ \cmjjj\ and a low-density phase with $n_\mathrm{H}\approx10^{-4}$ \cmjjj. 
With the remaining components,  our analysis finds a wide variation in gas densities, ranging from log\,$n_\mathrm{H}/ \cmjjj\sim-4$ to log\,$n_\mathrm{H}/ \cmjjj\sim-2$. Two of these components (component 6 and the lower-density phase of component 3) have low inferred densities of log\,$n_\mathrm{H}/ \cmjjj\approx-4$. Our \textsc{cloudy} photoionization models are able to reproduce the observed \ion{O}{VI} absorption in these two components. 

With the exception of one kinematic component, most of the gas associated with this LLS exhibits a moderate level of chemical enrichment, with inferred metallicities of between $\mathrm{[\alpha/H]}=-1.0\pm0.1$ and $\mathrm{[\alpha/H]}=-0.6\pm0.3$. In contrast, component 2 (which is optically thick) is significantly more metal poor with $\mathrm{[\alpha/H]}\approx-1.9$. The low metallicity of this component is consistent with its low $\mathrm{[N/\alpha]}=-0.4\pm0.2$, which indicates that the gas is likely to have been enriched at early times (see Figure 8). 

This LLS also exhibits a moderately strong absorption of high-ionization \ion{O}{VI}, which is accompanied by a weak \ion{N}{V} absorption. The \ion{O}{VI} absorption consists of a minimum of three components with a total log\,$N\mathrm{(\ion{O}{VI})/\cmjj}=14.5\pm0.1$. Two of the \ion{O}{VI} components are kinematically coincident with lower ionization gas and have narrow linewidths that are consistent with photoionized gas. Our photoionization models are able to reproduce the observed \ion{O}{VI} column densities in these components, as well as those of the lower ionization metal species. 

\begin{figure*}
\includegraphics[scale=1.4]{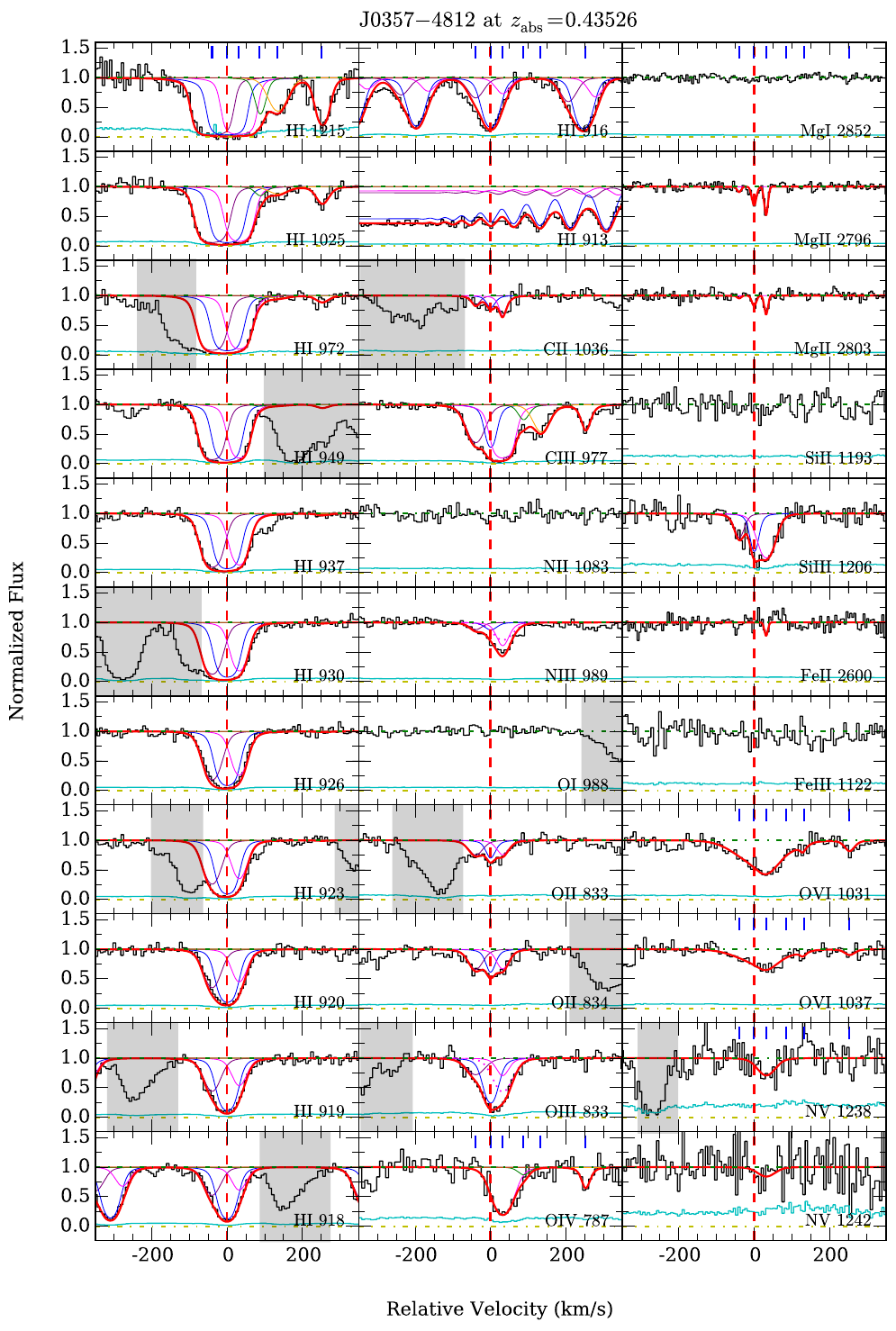}
\vspace{-0.75em}
\caption{Similar to Figure A1, but for the LLS at $z_\mathrm{abs}=0.43526$ toward J0357$-$4812. For the component at $dv_c=+32$ \kms, the two different gas phases required by our photoionization modeling are shown in dashed and solid magenta curves (for \ion{N}{III} and \ion{O}{III}) for the low-density and high-density gas phases, respectively (see text in Appendix A2).  }
\label{figure:ions}
\end{figure*}

\subsection{LLS at $z_\mathrm{abs}=0.62255$ toward J2135$-$5316}

This LLS is found at $z_\mathrm{abs}=0.62255$ and has a total $N\mathrm{(\ion{H}{I})}$ of log\,$N\mathrm{(\ion{H}{I})/\cmjj}=18.01\pm0.04$. The following metal species are also securely detected: \ion{C}{II}, \ion{C}{III}, \ion{N}{III}, \ion{O}{I},  \ion{Mg}{I}, \ion{Mg}{II}, \ion{Fe}{II}, and  \ion{O}{VI} (see Figure A3). We further note that while there is also possible \ion{N}{II} $\lambda 1083$ absorption, it is contaminated by a strong Ly$\alpha$ absorber at $z=0.4468$. For that reason, we exclude the possible \ion{N}{II} detection from our subsequent analysis. 

Our Voigt profile analysis identifies a minimum of six kinematic components in the absorption system, as shown in Figure A3. The observed kinematic spread of the absorber is $\Delta v\approx370$ \kms. Component 3 at $z_\mathrm{abs}=0.62255$ exhibits the strongest \ion{H}{I} absorption, comprising nearly all (98 percent) of the total $N\mathrm{(\ion{H}{I})}$. As shown in Figure A3, the component structure of \ion{H}{I} is in very good agreement with that of the metal ions. Note that no associated metal absorption is detected in components 1 and 6. The observed \ion{H}{I} linewidths in all components constrain the gas temperature to $T\lesssim2\times 10^5$ K. For components also detected in low-ionization species \ion{Mg}{II}, the observed ratios of the \ion{H}{I} and \ion{Mg}{II} linewidths are consistent with a cool gas with $T\approx 2\times10^4$ K and a moderate amount of non-thermal line broadening, $b_\mathrm{nt}\approx5-10$\,\kms.

This LLS exhibits a strong absorption of high-ionization \ion{O}{VI}. The \ion{O}{VI} absorption consists of two kinematic components with $b_c\approx(20-25)$ \kms\ and a total log\,$N\mathrm{(\ion{O}{VI})/\cmjj}=14.8\pm0.1$. The stronger \ion{O}{VI} component is coincident in velocity space with component 2 at $dv_c=-120$ \kms, which is also detected in \ion{H}{I} and \ion{C}{III}. The \ion{O}{VI} linewidth is comparable to both the \ion{H}{I} and \ion{C}{III} linewidths, which constrains the gas temperature to $T\lesssim 4\times10^4$ K and suggests a photoionized origin. While photoionization models utilizing the HM05 UVB radiation field can reproduce the observed \ion{C}{III} to \ion{O}{VI} column density ratio for a gas with log\,$n_\mathrm{H}/ \cmjjj\approx-4.4$, the implied absorber thickness is enormous,  $\ell>100$ kpc. Such a large size is comparable to the dimension of typical galaxy haloes and is difficult to reconcile with the narrow observed linewidths. For that reason, we consider this solution unphysical and seek an alternative explanation. 

We consider a scenario in which a nearby active galactic nucleus (AGN) provides additional ionizing photons to this absorption system. In Paper I, we noted the presence of a massive galaxy possibly hosting an active galactic nucleus (AGN) at $d\approx240$ kpc away from the QSO sightline. Based on the observed $\mathrm{[\ion{O}{III}]}\,5008$ emission flux of the host galaxy, we estimate that the AGN has a rest-frame NUV absolute magnitude of $M_\mathrm{UV}(3000)\approx -19.5$ (e.g., Reyes \etal\ 2008). Assuming that the AGN's NUV spectral energy distribution follows a power law described by $f_\nu \propto \nu^\alpha$, and adopting a reasonable range of spectral indices $\alpha=(-1.5,-0.5)$, we find that the added ionizing photons from an AGN at $r<500$ kpc lead to physically viable photoionization solutions as long as $\alpha>-1.5$. 

Similarly, the observed ionic ratios for the optically thick component 3 requires the addition of a local ionizing source to the extragalactic UVB, without which the column densities of low-ionization species (e.g., \ion{Mg}{ii}, \ion{Fe}{ii}) are overpredicted by $>0.5$ dex. For this component, we find that an AGN power spectrum with $\alpha\approx-0.5$ is required to reproduce the observed column densities of the low ions. Given these assumptions, the best-fitting ionizing model for component 3 is achieved for a gas density of log\,$n_\mathrm{H}/ \cmjjj\approx-0.7\pm0.2$ and a subsolar metallicity of $\mathrm{[\alpha/H]}\approx-0.9\pm0.1$. 

For the remaining components, we are able to reproduce the observed ionic column densities using our fiducial UVB without invoking additional ionizing photons. When compared to the need to invoke a nearby AGN to explain the observations in the other two components, this discrepancy can be understood to be due to (1) the gas being situated farther away from the AGN, (2) an anisotropic distribution of ionizing photons coming from the AGN (e.g., Lau \etal\ 2018), or (3) a combination of both effects. These components range from log\,$n_\mathrm{H}/ \cmjjj\approx-2$ (component 4) to log\,$n_\mathrm{H}/ \cmjjj\lesssim-3.6$ (component 5), with sub-solar metallicities ranging from $\mathrm{[\alpha/H]}\lesssim-0.3$ (component 5) to $\mathrm{[\alpha/H]}=-0.2\pm0.1$ (component 4). 

\begin{figure*}
\includegraphics[scale=1.3]{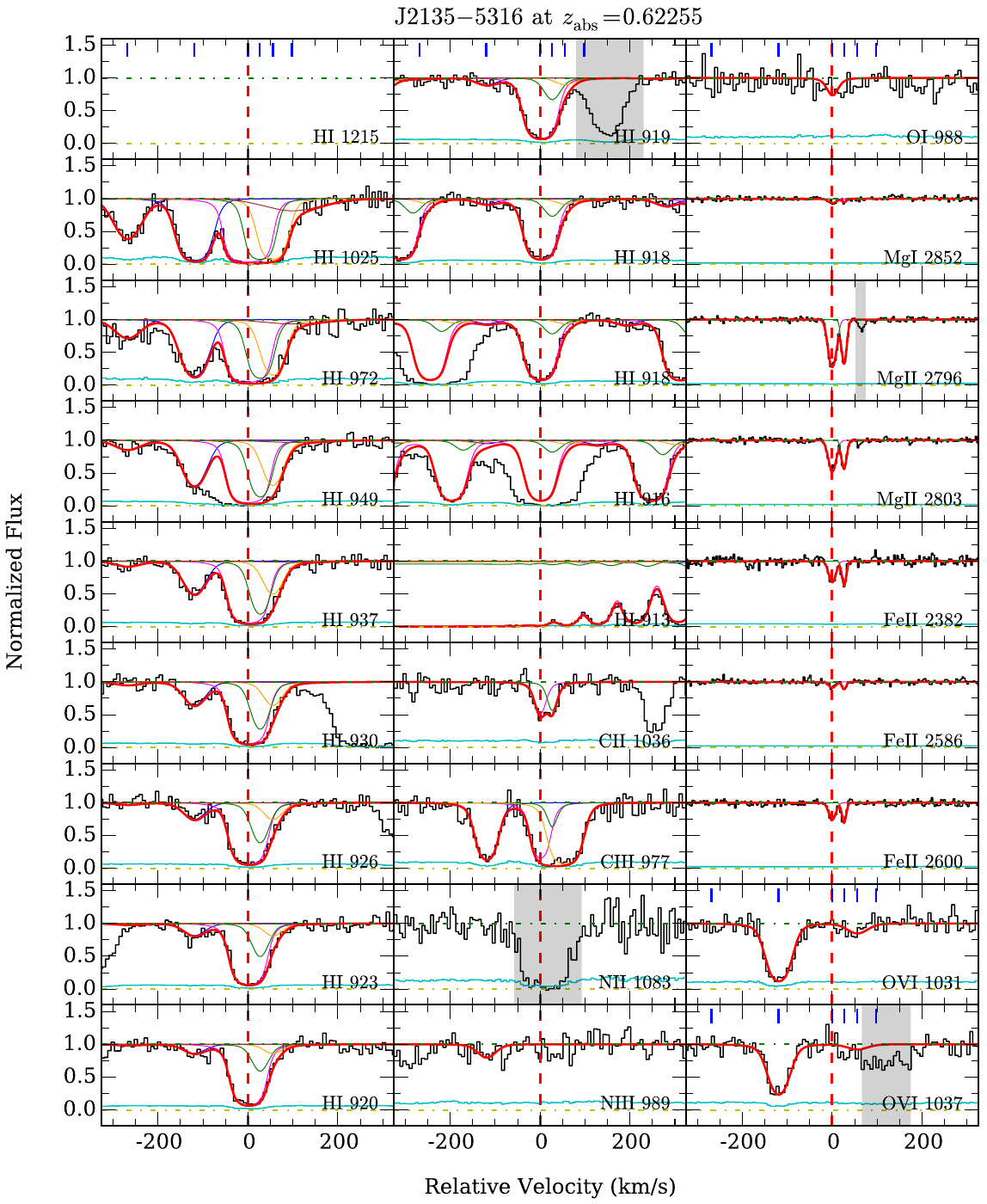}
\vspace{-0.75em}
\caption{Similar to Figure A1, but for the LLS at $z_\mathrm{abs}=0.62255$ toward J2135$-$5316.}
\label{figure:ions}
\end{figure*}

\subsection{LLS at $z_\mathrm{abs}=0.54273$ toward J2308$-$5258}

This LLS is situated at $z_\mathrm{abs}=0.54273$ and it has a total $N\mathrm{(\ion{H}{I})}$ of log\,$N\mathrm{(\ion{H}{I})/\cmjj}=17.59\pm0.02$. We detect associated absorption from the following metal species: \ion{C}{II}, \ion{C}{III}, \ion{N}{II},  \ion{O}{II},  \ion{Mg}{I}, \ion{Mg}{II}, \ion{S}{II}, \ion{S}{III}, and \ion{Fe}{II} (see Figure A4). While there is also possible \ion{N}{III} $\lambda 989$ absorption, the \ion{N}{III} $\lambda 989$ line is often contaminated by the adjacent \ion{Si}{II} $\lambda989$ line. Unfortunately, we cannot assess how much the \ion{N}{III} absorption is contaminated by \ion{Si}{II}, because no other \ion{Si}{II} transition is found within the spectral coverage of our data.  Therefore, we exclude this possible \ion{N}{III} detection from subsequent analysis. 

Our Voigt profile analysis identifies a minimum of six kinematic components, which is clear from Figure A4. The full kinematic spread of the absorber is $\Delta v\approx190$ \kms. Component 4 at $z_\mathrm{abs}=0.54273$ exhibits the strongest \ion{H}{I} absorption, comprising almost 75 percent of the total $N\mathrm{(\ion{H}{I})}$ of the system. The \ion{H}{I} kinematic structure is in excellent agreement with that of the metal ions. The observed \ion{H}{I} linewidths for all components constrain the gas temperature to $T\lesssim2\times 10^5$ K. For the five components also detected in \ion{Mg}{II}, the linewidth ratios of \ion{H}{I} and \ion{Mg}{II} in each component is indicates a cool gas temperature of $T\approx (1-2)\times10^4$ K, with modest non-thermal/bulk line broadening of $b_\mathrm{nt}\lesssim6$\,\kms.

Our ionization analysis indicates a moderate variation ($\sim0.5$ dex) in gas densities across the different components, with the exception of component 2 which exhibits the lowest \ion{H}{I} column density and a low inferred gas density of $n_\mathrm{H}/ \cmjjj\approx-3.5$. For the remaining  components, the observed ionic column densities are well-reproduced by models with densities ranging from log\,$n_\mathrm{H}/ \cmjjj\approx-2.3$ to log\,$n_\mathrm{H}/ \cmjjj\approx-1.4$ under the HM05 UVB. With the exception of component 2, which has a poorly constrained metallicity, the observed ionic column densities for all components indicate that the gas has been highly enriched in heavy elements. For the remaining five components, the inferred metallicities range from $\mathrm{[\alpha/H]}=-0.4\pm0.1$ (component 3) to $\mathrm{[\alpha/H]}=0.1\pm0.1$ (component 6). The inferred high metallicities are consistent with the observed column densities of C, N, and Fe ions, which require relative abundance ratios which are at least solar, $\mathrm{[C, N, Fe/\alpha]}\gtrsim 0.0$ to reproduce the observations. 

This LLS exhibits a strong absorption of high-ionization species \ion{O}{VI}. The \ion{O}{VI} absorption consists of a minimum of three broad components ($b_c\gtrsim30$ \kms) with a total log\,$N\mathrm{(\ion{O}{VI})/\cmjj}=14.8\pm0.1$. 

\begin{figure*}
\includegraphics[scale=1.4]{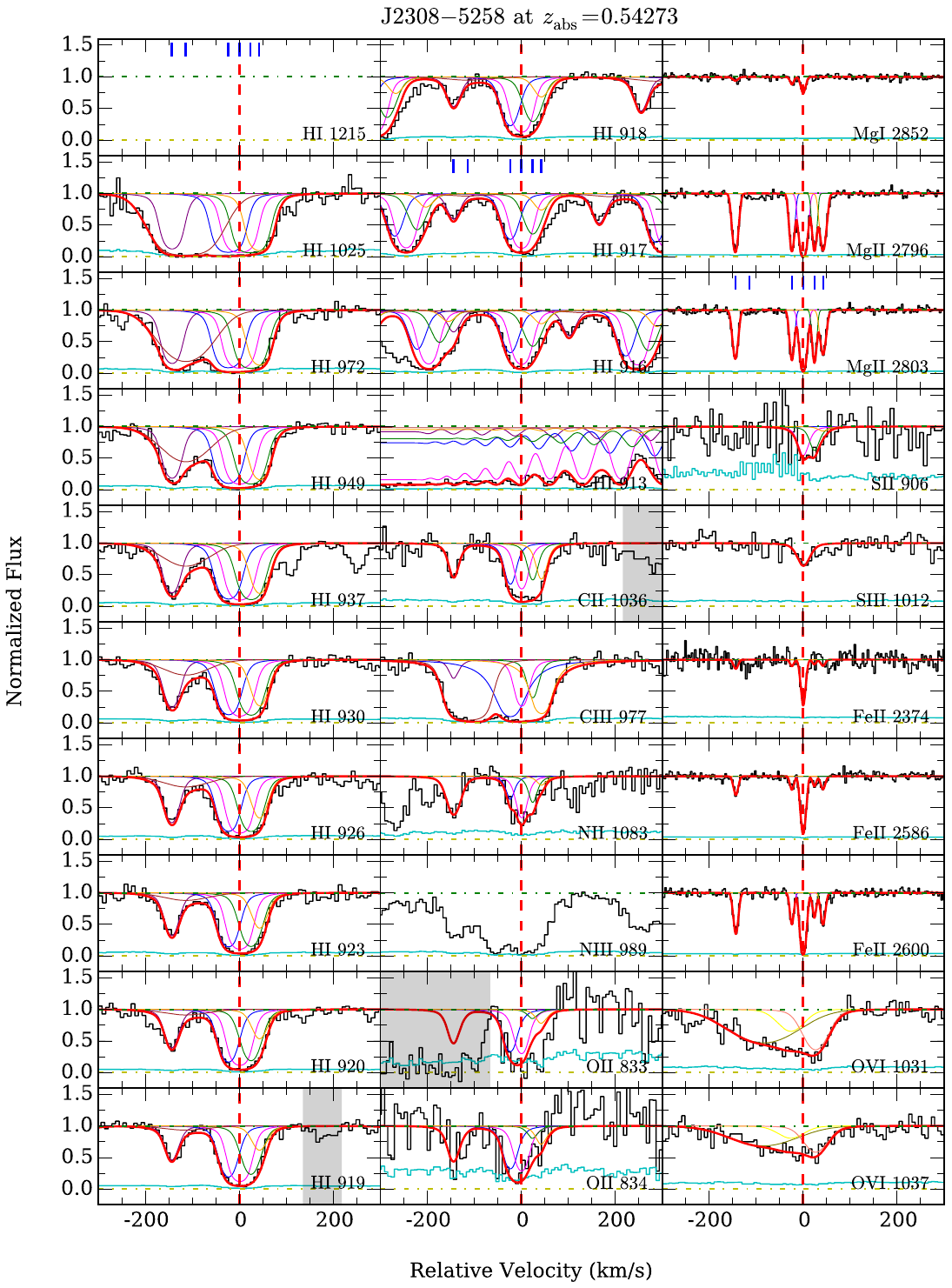}
\vspace{-0.75em}
\caption{Similar to Figure A1, but for the LLS at $z_\mathrm{abs}=0.54273$ toward J2308$-$5258. While there is possible \ion{N}{III} $\lambda 989$ absorption in this system, it is contaminated by the adjacent \ion{Si}{II} $\lambda989$ line.}
\label{figure:ions}
\end{figure*}

\begin{table}
\begin{center}
\begin{scriptsize}
\caption{Absorption properties for the LLS at $z_\mathrm{abs}=0.36400$ toward J0248$-$4048}
\hspace{-2.5em}
\vspace{-0.5em}
\label{}
\resizebox{2.8in}{!}{
\begin{tabular}{clrrr}\hline

Component	&	Species		&\multicolumn{1}{c}{$dv_c$} 		& \multicolumn{1}{c}{log\,$N_c/\cmjj$}	&\multicolumn{1}{c}{$b_c$}		\\	
 			&				&\multicolumn{1}{c}{(km\,s$^{-1}$)}	&   		   					& \multicolumn{1}{c}{(km\,s$^{-1}$)}  \\ \hline

1	& \ion{H}{I}	&	$-130.8^{+1.2}_{-1.0}$	& $15.42\pm0.02$		& $31.1^{+1.2}_{-0.8}$ \\  
	& \ion{C}{II}	&	$-130.8$				& $<12.6$				& 10	\\	
	& \ion{C}{III}	&	$-130.8$				& $13.89\pm0.04$		& $27.2^{+2.1}_{-1.8}$ 	\\
	& \ion{N}{II}	&	$-130.8$				& $<12.8$				& 10	\\
	& \ion{N}{III}	&	$-130.8$				& $<13.86\pm0.05$		& 31.1	\\
	& \ion{O}{I}	&	$-130.8$				& $<13.5$				& 10	\\
	& \ion{Si}{II}	&	$-130.8$				& $<12.4$				& 10	\\	
	& \ion{Si}{III}	&	$-130.8$				& $12.57\pm0.09$		& $22.2^{+10.1}_{-5.2}$	\\	
	& \ion{Mg}{I}	&	$-130.8$				& $<10.9$				& 10	\\	
	& \ion{Mg}{II}	&	$-130.8$				& $<11.2$				& 10	\\	
	& \ion{S}{II}	&	$-130.8$				& $<13.1$				& 10	\\	
	& \ion{S}{III}	&	$-130.8$				& $<13.6$				& 10	\\
	& \ion{Fe}{II}	&	$-130.8$				& $<12.1$				& 10	\\ 
	& \ion{Fe}{III}	&	$-130.8$				& $<12.9$				& 10	\\ \hline

2	&  \ion{H}{I}	&	$-42.1$				& $16.64\pm0.02$		& $23.8^{+0.5}_{-0.7}$  \\    
	& \ion{C}{II}	&	$-42.1$				& $13.67^{+0.14}_{-0.18}$& $8.7^{+4.1}_{-2.0}$	\\	
	& \ion{C}{III}	&	$-42.1$				& $>14.43$				& $<40.7$ 	\\
	& \ion{N}{II}	&	$-42.1$				& $<12.8$			 	& 10 \\
	& \ion{N}{III}	&	$-42.1$				& $14.08\pm0.05$		& 23.8	\\
	& \ion{O}{I}	&	$-42.1$				& $<13.5$				& 10	\\
	& \ion{Si}{II}	&	$-42.1$				& $12.53^{+0.11}_{-0.19}$	& $8.0^{+2.9}_{-2.5}$	\\	
	& \ion{Si}{III}	&	$-42.1$				& $13.35^{+0.21}_{-0.08}$& $25.2^{+3.2}_{-5.7}$	\\
	& \ion{Mg}{I}	&	$-42.1$				& $<10.9$				& 10	\\	
	& \ion{Mg}{II}	&	$-42.1\pm0.5$			& $12.55\pm0.03$		& $8.7\pm0.7$  \\  
	& \ion{S}{II}	&	$-42.1$				& $<13.2$				& 10	\\	
	& \ion{S}{III}	&	$-42.1$				& $<13.6$				& 10	\\	
	& \ion{Fe}{II}	&	$-42.1$				& $12.46^{+0.11}_{-0.16}$& $6.1^{+4.0}_{-2.0}$	\\ 
	& \ion{Fe}{III}	&	$-42.1$				& $<12.9$				& 10		\\ \hline

3	&  \ion{H}{I}	&	$0.0$				& $17.51\pm0.01$		& $20.0\pm0.3$  \\  
	& \ion{C}{II}	&	$0.0$				& $14.35^{+0.07}_{-0.04}$& $20.1^{+7.6}_{-4.1}$ \\		
	& \ion{C}{III}	&	$0.0$				& $>14.74$			& $<31.4$ \\
	& \ion{N}{II}	&	$0.0$				& $14.06^{+0.11}_{-0.06}$	& $11.5^{+4.1}_{-2.6}$	\\
	& \ion{N}{III}	&	$0.0$				& $14.17^{+0.12}_{-0.06}$& $13.9^{+3.8}_{-3.7}$	\\
	& \ion{O}{I}	&	$0.0$				& $13.80^{+0.10}_{-0.19}$& $12.0^{+2.6}_{-4.8}$	\\
	& \ion{Si}{II}	&	$0.0$				& $13.76^{+0.12}_{-0.10}$& $10.0^{+1.5}_{-1.2}$ \\	
	& \ion{Si}{III}	&	$0.0$				& $>13.54$			& $<27.2$ \\	
	& \ion{Mg}{I}	&	$0.0$				& $11.73^{+0.09}_{-0.04}$	& $12.4\pm2.0$ \\	
	& \ion{Mg}{II}	&	$0.0\pm0.2$			& $13.44\pm0.03$		& $10.9\pm0.3$  \\ 
	& \ion{S}{II}	&	$0.0$				& $<13.1$				& 10 \\  
	& \ion{S}{III}	&	$0.0$				& $14.09\pm0.14$		& $10^{+3.0}_{-3.5}$ \\  
	& \ion{Fe}{II}	&	$0.0$				& $13.46^{+0.03}_{-0.05}$& $10.8\pm0.8$	\\ 
	& \ion{Fe}{III}	&	$0.0$				& $14.04\pm0.08$		& $27.8^{+6.3}_{-5.5}$	\\ \hline

4	&  \ion{H}{I}	&	$+57.3^{+1.2}_{-1.6}$ 	& $15.36\pm0.04$		& $9.3^{+1.5}_{-1.0}$   \\   
	&\ion{C}{II}	&	$+57.3$				& $<14.03\pm0.06$		& $17.1^{+3.8}_{-3.5}$ \\		
	& \ion{C}{III}	&	$+57.3$				& $<13.46$			& $10$ 	\\ \\
	& \ion{N}{II}	&	$+57.3$				& $<12.8$				& 10	\\
	& \ion{N}{III}	&	$+57.3$				& $<12.9$				& 10	\\
	& \ion{O}{I}	&	$+57.3$				& $<13.5$			& 10	\\
	& \ion{Si}{II}	&	$+57.3$				& $<12.4$				& 10\\	
	& \ion{Si}{III}	&	$+57.3$				& $12.81^{+0.26}_{-0.15}$& $9.0\pm2.7$ \\	
	& \ion{Mg}{I}	&	$+57.3$				& $<10.9$				& $10$ \\	
	& \ion{Mg}{II}	&	$+57.3$				& $<11.2$				& $10$ \\	
	& \ion{Si}{II}	&	$+57.3$				& $<13.1$				& 10\\	
	& \ion{Si}{III}	&	$+57.3$				& $<13.6$				& $10$ \\	
	& \ion{Fe}{II}	&	$+57.3$				& $<12.1$				& 10	\\ 
	& \ion{Fe}{III}	&	$+57.3$				& $<12.9$				& 10	\\ \hline

5	&  \ion{H}{I}	&	$+100.0^{+8.4}_{-10.1}$ 	& $14.70\pm0.08$		& $69.0^{+6.1}_{-5.6}$  \\  
	& \ion{C}{II}	&	$+100.0$				& $<12.6$				& 10 \\		
	& \ion{C}{III}	&	$+100.0$				& $13.43^{+0.04}_{-0.08}$& $69$ \\
	& \ion{N}{II}	&	$+100.0$				& $<12.8$				& 10	\\
	& \ion{N}{III}	&	$+100.0$				& $<12.9$				& 10	\\
	& \ion{O}{I}	&	$+100.0$				& $<13.5 $			& 10	\\
	& \ion{Si}{II}	&	$+100.0$				& $<12.4	$			& 10 \\	
	& \ion{Si}{III}	&	$+100.0$				& $<12.6$				& $10$ \\	
	& \ion{Mg}{I}	&	$+100.0$				& $<10.9$				& 10 \\	
	& \ion{Mg}{II}	&	$+100.0$				& $<11.2$				& $10$ \\	
	& \ion{S}{II}	&	$+100.0$				& $<13.1$				& 10 \\	
	& \ion{S}{III}	&	$+100.0$				& $<13.6$				& $10$ \\		
	& \ion{Fe}{II}	&	$+100.0$				& $<12.1$				& 10	\\ 
	& \ion{Fe}{III}	&	$+100.0$				& $<12.9$				& 10	\\ \hline

6	& \ion{H}{I}	&	$+154.4^{+1.5}_{-1.6}$ 	& $14.59\pm0.05$		& $19.2^{+2.3}_{-1.7}$  \\  \
	& \ion{C}{III}	&	$+154.4$				& $13.56\pm0.04$		& $26.8^{+3.5}_{-3.0}$ \\
	& \ion{N}{II}	&	$+154.4$				& $<12.8$				& 10	\\
	& \ion{N}{III}	&	$+154.4$				& $<12.9$				& 10	\\
	& \ion{O}{I}	&	$+154.4$				& $<	13.5	$			& 10	\\
	& \ion{Si}{II}	&	$+154.4$				& $<12.4$				& 10 \\	
	& \ion{Si}{III}	&	$+154.4$				& $12.11^{+0.11}_{-0.32}$	& $19.2$ \\		
	& \ion{Mg}{I}	&	$+154.4$				& $<10.9$				& 10 \\	
	& \ion{Mg}{II}	&	$+154.4$				& $<11.2$				& $10$ \\		
	& \ion{S}{II}	&	$+154.4$				& $<13.2$				& $10$ \\		
	& \ion{S}{III}	&	$+154.4$				& $<13.6$				& 10	\\ 
	& \ion{Fe}{II}	&	$+154.4$				& $<12.1$				& 10	\\ 
	& \ion{Fe}{III}	&	$+154.4$				& $<12.9$				& 10	\\  \hline

high-1	& \ion{O}{VI}	&	$-121.9\pm5.1$		& $14.05\pm0.06$			& $42.9\pm7.8$\\ 
		& \ion{N}{V}	&	$-121.9$			& $<13.2$					& $42.9$\\ 
high-2	& \ion{O}{VI}	&	$-12.8\pm2.9$		& $14.42\pm0.04$			& $41.2\pm6.0$\\  
		& \ion{N}{V}	&	$-12.8$			& $<13.2$					& $41.2$\\ 
high-3	& \ion{O}{VI}	&	$+57.2\pm6.5$		& $13.52\pm0.29$			& $15.5\pm13.3$\\  
		& \ion{N}{V}	&	$+57.2$			& $<13.0$					& $15.5$\\ 
high-4	& \ion{O}{VI}	&	$+134.6\pm12.4$	& $14.21\pm0.07$			& $82.6\pm16.0$\\ 
		& \ion{N}{V}	&	$+134.6$			& $<13.4$				& $82.6$\\ 

\hline
\end{tabular}}
\end{scriptsize}
\end{center}
\end{table}

\begin{table}
\begin{center}
\begin{tiny}
\caption{Absorption properties for the LLS at $z_\mathrm{abs}=0.43526$ toward J0357$-$4812}
\hspace{-2.5em}
\vspace{-0.5em}
\label{}
\resizebox{2.9in}{!}{
\begin{tabular}{clrrr}\hline

Component	&	Species		&\multicolumn{1}{c}{$dv_c$} 		& \multicolumn{1}{c}{log\,$N_c/\cmjj$}	&\multicolumn{1}{c}{$b_c$}		\\	
 			&				&\multicolumn{1}{c}{(km\,s$^{-1}$)}	&   		   					& \multicolumn{1}{c}{(km\,s$^{-1}$)}  \\ \hline 

1	& \ion{H}{I}	&	$-39.9$				& $16.23\pm0.02$				& $22.2^{+0.4}_{-0.3}$ \\  
	& \ion{C}{II}	&	$-39.9$				& $13.28^{+0.12}_{-0.18}$		& $47.1^{+3.9}_{-0.9}$	\\	
	& \ion{C}{III}	&	$-39.9$				& $13.55^{+0.04}_{-0.03}$		& $24.3^{+2.4}_{-3.0}$ 	\\
	& \ion{N}{II}	&	$-39.9$				& $<12.9$						& 10	\\
	& \ion{N}{III}	&	$-39.9$				& $13.35^{+0.10}_{-0.13}$		& $16.5^{+3.3}_{-2.7}$	\\
	& \ion{O}{I}	&	$-39.9$				& $<13.2$						& 10	\\
	& \ion{O}{II}	&	$-39.9$				& $13.89^{+0.04}_{-0.07}$		& $19.1^{+1.4}_{-6.2}$	\\
	& \ion{O}{III}	&	$-39.9$				& $13.95^{+0.05}_{-0.14}$		& $26.8^{+2.1}_{-5.5}$	\\
	& \ion{O}{IV}	&	$-39.9$				& $<13.3$						& 10	\\
	& \ion{Mg}{I}	&	$-39.9$				& $<10.8$						& 10	\\	
	& \ion{Mg}{II}	&	$-39.9\pm3.0$			& $11.47\pm0.16$				& $5$	\\	
	& \ion{Si}{II}	&	$-39.9$				& $<12.4$						& 10	\\	
	& \ion{Si}{III}	&	$-39.9$				& $12.59^{+0.13}_{-0.12}$		& $12.0^{+5.3}_{-4.9}$	\\	
	& \ion{Fe}{II}	&	$-39.9$				& $<12.1$						& 10	\\ 
	& \ion{Fe}{III}	&	$-39.3$				& $<13.2$				& 10	\\  \hline

2	&  \ion{H}{I}	&	$0.0$				& $17.09\pm0.01$				& $18.6\pm0.6$  \\  
	& \ion{C}{II}	&	$0.0$				& $13.42^{+0.08}_{-0.21}$		& $7.9^{+2.8}_{-2.5}$	\\	
	& \ion{C}{III}	&	$0.0$				& $>13.88$					& $<9.9$ 	\\
	& \ion{N}{II}	&	$0.0$				& $<12.9$						& 10	\\
	& \ion{N}{III}	&	$0.0$				& $13.44^{+0.06}_{-0.29}$		& $16.0^{+1.6}_{-6.1}$	\\
	& \ion{O}{I}	&	$0.0$				& $<13.1$						& 10	\\
	& \ion{O}{II}	&	$0.0$				& $13.90^{+0.12}_{-0.05}$		& $9.1^{+5.8}_{-1.6}$	\\
	& \ion{O}{III}	&	$0.0$				& $14.68^{+0.27}_{-0.09}$		& $15.9^{+3.6}_{-3.2}$	\\
	& \ion{O}{IV}	&	$0.0$				& $<13.3$						& 10	\\
	& \ion{Mg}{I}	&	$0.0$				& $<10.8$						& 10	\\	
	&  \ion{Mg}{II}	&	$0.0\pm0.8$			& $12.10\pm0.05$				& $5.6\pm1.7$  \\  
	& \ion{Si}{II}	&	$0.0$				& $<12.4$						& 10	\\	
	& \ion{Si}{III}	&	$0.0$				& $>12.75$					& $<15.1$	\\
	& \ion{Fe}{II}	&	$0.0$				& $<12.1$						& 10	\\ 
	& \ion{Fe}{III}	&	$0.0$				& $<13.2$				& 10	\\  \hline
	
3	&  \ion{H}{I}	&	$+31.8$				& $16.07\pm0.03$			& $19.3\pm0.6$  \\  
	& \ion{C}{II}	&	$+31.8$				& $13.89^{+0.02}_{-0.28}$	& $4.2$ \\		
	& \ion{C}{III}	&	$+31.8$				& $>14.64$				& $<16.6$ \\
	& \ion{N}{II}	&	$+31.8$				& $<12.9$					& 10	\\
	& \ion{N}{III}	&	$+31.8$				& $13.74^{+0.09}_{-0.05}$	& $14.8$	\\
	& \ion{N}{III}	&	$+31.8$				& $14.06^{+0.06}_{-0.06}$	& $27.9$	\\
	& \ion{O}{I}	&	$+31.8$				& $<13.2$					& 10	\\
	& \ion{O}{II}	&	$+31.8$				& $13.83\pm0.08$		& $14.8^{+3.0}_{-6.7}$	\\
	& \ion{O}{III}	&	$+31.8$				& $13.93^{+0.31}_{-0.42}$	& $14.8$	\\
	& \ion{O}{III}	&	$+31.8$				& $14.37^{+0.08}_{-0.15}$	& $27.9$	\\
	& \ion{O}{IV}	&	$+31.8$				& $14.77^{+0.19}_{-0.06}$	& $27.9^{+5.5}_{-5.4}$	\\
	& \ion{Mg}{I}	&	$+31.8$				& $<10.8$				& 10 \\	
	& \ion{Mg}{II}	&	$+31.8\pm0.6$			& $12.38\pm0.11$		& $2.9\pm1.0$  \\  
	& \ion{Si}{II}	&	$+31.8$				& $<12.4$				& 10 \\	
	& \ion{Si}{III}	&	$+31.8$				& $13.22^{+0.15}_{-0.06}$& $23.1^{+3.1}_{-5.8}$ \\	
	& \ion{Fe}{II}	&	$+31.8$				& $12.37\pm0.37$		& $2.1\pm6.3$	\\ 
	& \ion{Fe}{III}	&	$+31.8$				& $<13.2$				& 10	\\  \hline

4	&  \ion{H}{I}	&	$+87.8\pm7.7$			& $13.52^{+0.10}_{-0.14}$& $12.9\pm2.8$  \\  
	& \ion{C}{II}	&	$+87.8$				& $<12.8$				& 10 \\		
	& \ion{C}{III}	&	$+87.8$				& $12.84^{+0.05}_{-0.13}$& $13.0\pm0.8$ \\
	& \ion{N}{II}	&	$+87.8$				& $<12.9$				& 10	\\
	& \ion{N}{III}	&	$+87.8$				& $<12.7$				& 10	\\
	& \ion{O}{I}	&	$+87.8$				& $<13.2$				& 10	\\
	& \ion{O}{II}	&	$+87.8$				& $<12.8	$			& 10	\\
	& \ion{O}{III}	&	$+87.8$				& $<13.0	$			& 10	\\
	& \ion{O}{IV}	&	$+87.8$				& $<13.68$			& $<18.5$	\\
	& \ion{Mg}{I}	&	$+87.8$				& $<10.8$				& $10$ \\	
	& \ion{Mg}{II}	&	$+87.8$				& $<11.3$				& $10$ \\	
	& \ion{Si}{II}	&	$+87.8$				& $<12.4$				& 10\\	
	& \ion{Si}{III}	&	$+87.8$				& $<11.8$				& $10$ \\	
	& \ion{Fe}{II}	&	$+87.8$				& $<12.1$				& 10	\\ 
	& \ion{Fe}{III}	&	$+87.8$				& $<13.2$				& 10	\\ \hline

5	&  \ion{H}{I}	&	$+134.3\pm8.3$		& $13.60^{+0.07}_{-0.06}$		& $27.9^{+3.7}_{-2.8}$  \\  
	& \ion{C}{II}	&	$+134.3$				& $<12.8$						& 10 \\		
	& \ion{C}{III}	&	$+134.3$				& $13.28^{+0.03}_{-0.04}$		& $22.0^{+2.2}_{-2.0}$ \\
	& \ion{N}{II}	&	$+134.3$				& $<12.9$				& 10	\\
	& \ion{N}{III}	&	$+134.3$				& $<12.7$				& 10	\\
	& \ion{O}{II}	&	$+134.3$				& $<13.2 $			& 10	\\
	& \ion{O}{II}	&	$+134.3$				& $<12.8 $			& 10	\\
	& \ion{O}{III}	&	$+134.3$				& $<	13.0	$			& 10	\\
	& \ion{O}{IV}	&	$+134.3$				& $<13.3	$			& 10	\\
	& \ion{Mg}{I}	&	$+134.3$				& $<10.8$				& 10 \\	
	& \ion{Mg}{II}	&	$+134.3$				& $<11.3$				& $10$ \\		
	& \ion{Si}{II}	&	$+134.3$				& $<12.4$				& 10 \\	
	& \ion{Si}{III}	&	$+134.3$				& $<11.8$				& $10$ \\	
	& \ion{Fe}{II}	&	$+134.3$				& $<12.1$				& 10	\\ 
	& \ion{Fe}{III}	&	$+134.3$				& $<13.2$				& 10	\\ \hline

6	& \ion{H}{I}	&	$+252.9^{+2.1}_{-1.7}$	& $13.89^{+0.07}_{-0.06}$& $16.2^{+2.1}_{-2.0}$  \\  
	& \ion{C}{II}	&	$+252.9$				& $<12.8$				& 10 \\		
	& \ion{C}{III}	&	$+252.9$				& $13.18^{+0.15}_{-0.07}$& $8.6^{+1.5}_{-2.3}$ \\
	& \ion{N}{II}	&	$+252.9$				& $<12.9$				& 10	\\
	& \ion{N}{III}	&	$+252.9$				& $<12.7$				& 10	\\
	& \ion{O}{I}	&	$+252.9$				& $<	13.2	$			& 10	\\
	& \ion{O}{II}	&	$+252.9$				& $<	12.8	$			& 10	\\
	& \ion{O}{III}	&	$+252.9$				& $<13.0$				& 10	\\
	& \ion{O}{IV}	&	$+252.9$				& $13.86^{+0.13}_{-0.17}$& $10.9^{+4.2}_{-2.2}$\\
	& \ion{Mg}{I}	&	$+252.9$				& $<10.8$				& 10 \\	
	& \ion{Mg}{II}	&	$+252.9$				& $<11.3$				& $10$ \\		
	& \ion{Si}{II}	&	$+252.9$				& $<12.4$				& 10 \\	
	& \ion{Si}{III}	&	$+252.9$				& $<11.8$				& $10$ \\		
	& \ion{Fe}{II}	&	$+252.9$				& $<12.1$				& 10	\\ 
	& \ion{Fe}{III}	&	$+252.9$				& $<13.2$				& 10	\\  \hline
high-1	& \ion{O}{VI}	&	$+18.3\pm4.0$			& $14.33\pm0.04$		& $77.1\pm7.2$	\\
		& \ion{N}{V}	&	$+18.3$				& $<13.4$				& $77.1$	\\
high-2	& \ion{O}{VI}	&	$+31.8$				& $14.05\pm0.06$		& $27.9$	\\
		& \ion{N}{V}	&	$+31.8$				& $13.61\pm0.14$		& $27.9$	\\
high-3	& \ion{O}{VI}	&	$+252.9$				& $13.44^{+0.10}_{-0.12}$& $13.9^{+3.0}_{-7.7}$\\
		& \ion{N}{V}	&	$+252.9$				& $<13.1$				& $13.9$\\

\hline
\end{tabular}}
\end{tiny}
\end{center}
\end{table}

\begin{table}
\begin{center}
\begin{scriptsize}
\caption{Absorption properties for the LLS at $z_\mathrm{abs}=0.62255$ toward J2135$-$5316}
\hspace{-2.5em}
\vspace{-0.5em}
\label{}
\resizebox{3.5in}{!}{
\begin{tabular}{clrrr}\hline

Component	&	Species		&\multicolumn{1}{c}{$dv_c$} 		& \multicolumn{1}{c}{log\,$N_c/\cmjj$}	&\multicolumn{1}{c}{$b_c$}		\\	
 			&				&\multicolumn{1}{c}{(km\,s$^{-1}$)}	&   		   					& \multicolumn{1}{c}{(km\,s$^{-1}$)}  \\ \hline 

1	& \ion{H}{I}	&	$-268.3^{+2.3}_{-2.6}$	& $14.51^{+0.03}_{-0.04}$& $32.5^{+2.7}_{-2.6}$ \\  
	& \ion{C}{II}	&	$-268.3$				& $<13.0$		& 10\\	
	& \ion{C}{III}	&	$-268.3$				& $<12.2$		& 10\\
	& \ion{N}{II}	&	$-268.3$				& $<13.1$		& 10\\
	& \ion{N}{III}	&	$-268.3$				& $<13.0$		& 10\\
	& \ion{O}{I}	&	$-268.3$				& $<13.4$		& 10\\
	& \ion{Mg}{I}	&	$-268.3$				& $<10.5$		& 10\\
	& \ion{Mg}{II}	&	$-268.3$				& $<11.0$		& 10\\
	& \ion{Fe}{II}	&	$-268.3$				& $<11.6$		& 10\\  \hline

2	&  \ion{H}{I}	&	$-119.7^{+0.5}_{-0.7}$	& $15.38\pm0.02$			& $26.8\pm0.6$  \\  
	& \ion{C}{II}	&	$-119.7$				& $<13.0$					& 10\\	
	& \ion{C}{III}	&	$-119.7$				& $14.24^{+0.36}_{-0.27}$	& $16.2^{+2.7}_{-2.6}$\\
	& \ion{N}{II}	&	$-119.7$				& $<13.1$					& 10\\
	& \ion{N}{III}	&	$-119.7$				& $13.65^{+0.10}_{-0.17}$	& $22.0^{+4.0}_{-3.7}$\\
	& \ion{O}{I}	&	$-119.7$				& $<13.4$					& 10\\
	& \ion{Mg}{I}	&	$-119.7$				& $<10.5$					& 10\\
	& \ion{Mg}{II}	&	$-119.7$				& $<11.0$					& 10\\
	& \ion{Fe}{II}	&	$-119.7$				& $<11.6$					& 10\\  \hline
	
3	& \ion{H}{I}	&	$0.0$				& $18.00\pm0.04$	& $18.7\pm0.2$  \\  
	& \ion{C}{II}	&	$0.0$				& $13.99^{+0.19}_{-0.15}$	& $8.1^{+4.6}_{-1.9}$\\	
	& \ion{C}{III}	&	$0.0$				& $>13.76$				& $<17.6$\\
	& \ion{N}{III}	&	$0.0$				& $<13.0$					& 10\\
	& \ion{O}{I}	&	$0.0$				& $13.99^{+0.14}_{-0.16}$	& 10.4\\
	& \ion{Mg}{I}	&	$0.0$				& $11.04\pm0.09$	& $11.2\pm3.2$\\
	& \ion{Mg}{II}	&	$0.0\pm0.2$			& $12.76\pm0.01$	& $10.4\pm0.3$\\
	& \ion{Fe}{II}	&	$0.0$				& $12.59\pm0.02$	& $9.6\pm0.8$ \\  \hline

4	& \ion{H}{I}	&	$+26.4$				& $15.87^{+0.10}_{-0.38}	$& $18.5^{+3.7}_{-2.8}$  \\  
	& \ion{C}{II}	&	$+26.4$				& $13.88^{+0.11}_{-0.16}$	& $7.7\pm5.9$\\	
	& \ion{C}{III}	&	$+26.4$				& $>13.28$		& $<14.1$\\
	& \ion{N}{III}	&	$+26.4$				& $<13.0$				& 10\\
	& \ion{O}{I}	&	$+26.4$				& $<13.4$				& $10$\\
	& \ion{Mg}{I}	&	$+26.4$				& $10.70\pm0.13$		& $5.8$\\
	& \ion{Mg}{II}	&	$+26.4\pm0.2$			& $12.53\pm0.01$		& $5.8\pm0.3$\\
	& \ion{Fe}{II}	&	$+26.4$				& $12.47\pm0.03$		& $4.5\pm0.7$ \\  \hline

5	&  \ion{H}{I}	&	$+55.5^{+3.6}_{-8.4}$	& $15.32^{+0.22}_{-0.11}$	& $21.9^{+4.5}_{-1.8}$  \\  
	& \ion{C}{II}	&	$+55.5$				& $<13.0$			& 10\\	
	& \ion{C}{III}	&	$+55.5$				& $>14.24$		& $<26.5$\\
	& \ion{N}{II}	&	$+55.5$				& $<13.1$			& 10\\
	& \ion{N}{III}	&	$+55.5$				& $<13.0$			& 10\\
	& \ion{O}{I}	&	$+55.5$				& $<13.4$			& 10\\
	& \ion{Mg}{I}	&	$+55.5$				& $<10.5$			& 10\\
	& \ion{Mg}{II}	&	$+55.5$				& $<11.0$			& 10\\
	& \ion{Fe}{II}	&	$+55.5$				& $<11.6$			& 10\\  \hline

6	& \ion{H}{I}	&	$+98.0^{+14.3}_{-13.4}$	& $14.17^{+0.10}_{-0.16}$& $83.0^{+17.9}_{-14.3}$  \\   
	& \ion{C}{II}	&	$+98.0$				& $<13.0$				& 10\\	
	& \ion{C}{III}	&	$+98.0$				& $<12.2$				& 10 \\
	& \ion{N}{II}	&	$+98.0$				& $<13.1$				& 10\\
	& \ion{N}{III}	&	$+98.0$				& $<13.0$				& 10\\
	& \ion{O}{I}	&	$+98.0$				& $<13.4$				& 10\\
	& \ion{Mg}{I}	&	$+98.0$				& $<10.5$				& 10\\
	& \ion{Mg}{II}	&	$+98.0$				& $<11.0$				& 10\\
	& \ion{Fe}{II}	&	$+98.0$				& $<11.6$				& 10\\  \hline
high-1	& \ion{O}{VI}	&	$-119.7$				& $14.73^{+0.09}_{-0.05}$	& $22.4^{+1.5}_{-2.0}$\\
high-2	& \ion{O}{VI}	&	$+55.5$				& $13.43^{+0.08}_{-0.25}$	& $25.9^{+14.3}_{-10.6}$\\

\hline
\end{tabular}}
\end{scriptsize}
\end{center}
\end{table}

\begin{table}
\begin{center}
\begin{scriptsize}
\caption{Absorption properties for the LLS at $z_\mathrm{abs}=0.54273$ toward J2308$-$5258}
\hspace{-2.5em}
\vspace{-0.5em}
\label{}
\resizebox{3.5in}{!}{
\begin{tabular}{clrrr}\hline

Component	&	Species		&\multicolumn{1}{c}{$dv_c$} 		& \multicolumn{1}{c}{log\,$N_c/\cmjj$}	&\multicolumn{1}{c}{$b_c$}		\\	
 			&				&\multicolumn{1}{c}{(km\,s$^{-1}$)}	&   		   					& \multicolumn{1}{c}{(km\,s$^{-1}$)}  \\ \hline 

1	& \ion{H}{I}	&	$-144.4$				& $16.12\pm0.03$	& $12.2^{+0.8}_{-0.6}$ \\  
	& \ion{C}{II}	&	$-144.4$				& $14.24^{+0.08}_{-0.33}$	& $6^{+3.6}_{-1.8}$\\	
	& \ion{C}{III}	&	$-144.4$				& $>13.05$				& $<16.8$\\
	& \ion{N}{II}	&	$-144.4$				& $14.20^{+0.16}_{-0.09}$	& $12.9\pm4.5$\\
	& \ion{O}{II}	&	$-144.4$				& $>13.98$				& $<9.5$\\
	& \ion{Mg}{I}	&	$-144.4$				& $10.82\pm0.17$			& $6.8\pm4.5$\\
	& \ion{Mg}{II}	&	$-144.4\pm0.1$			& $12.96\pm0.02$			& $5.8\pm0.2$\\
	& \ion{S}{II}	&	$-144.4$				& $<13.3$					& $10$\\
	& \ion{S}{III}	&	$-144.4$				& $<13.3$					& $10$\\
	& \ion{Fe}{II}	&	$-144.4$				& $12.97\pm0.02$			& $5.0\pm0.4$\\  \hline

2	&  \ion{H}{I}	&	$-114.3\pm3.1$			& $15.44\pm0.02$	&$59.6^{+3.3}_{-3.0}$  \\  
	& \ion{C}{II}	&	$-114.3$				& $<12.9$			& 10\\	
	& \ion{C}{III}	&	$-114.3$				& $>14.78$		& 	$<26.4$\\
	& \ion{N}{II}	&	$-114.3$				& $<13.0$			& 10\\
	& \ion{O}{II}	&	$-114.3$				& $<13.3$			& 10\\
	& \ion{Mg}{I}	&	$-114.3$				& $<10.7$			& 10\\
	& \ion{Mg}{II}	&	$-114.3$				& $<11.2$			& 10\\
	& \ion{S}{II}	&	$-114.3$				& $<13.4$			& 10\\
	& \ion{S}{III}	&	$-114.3$				& $<13.4$			& 10\\
	& \ion{Fe}{II}	&	$-114.3$				& $<11.8$			& 10\\  \hline
	
3	& \ion{H}{I}	&	$-23.9$				& $16.67^{+0.03}_{-0.05}$& $16.1^{+0.6}_{-0.5}$  \\  
	& \ion{C}{II}	&	$-23.9$				& $>14.06$			& $<10.2$\\	
	& \ion{C}{III}	&	$-23.9$				& $>14.24$			& $<19.9$\\
	& \ion{N}{II}	&	$-23.9$				& $13.83^{+0.23}_{-0.33}$& $5.0^{+1.3}_{-0.7}$\\
	& \ion{O}{II}	&	$-23.9$				& $>14.51$		& $<9.4$\\
	& \ion{Mg}{I}	&	$-23.9$				& $10.97 \pm0.10$		& $3.3\pm2.7$\\
	& \ion{Mg}{II}	&	$-23.9\pm0.1$			& $12.98\pm0.04$		& $4.4\pm0.2$\\
	& \ion{S}{II}	&	$-23.9$				& $<13.4$				& $10$\\
	& \ion{S}{III}	&	$-23.9$				& $<13.3$				& $10$\\
	& \ion{Fe}{II}	&	$-23.9$				& $12.73\pm0.04$		& $3.4\pm0.6$ \\  \hline

4	& \ion{H}{I}	&	$0.0$				& $17.46\pm0.02$		& $14.4^{+1.0}_{-0.7}$  \\  
	& \ion{C}{II}	&	$0.0$				& $>15.46$		& $<9.1$\\	
	& \ion{C}{III}	&	$0.0$				& $>13.47$		& $<19.6$\\
	& \ion{N}{II}	&	$0.0$				& $>14.13$		& $<11.1$\\
	& \ion{O}{II}	&	$0.0$				& $>14.83$		& $<10.9$\\
	& \ion{Mg}{I}	&	$0.0$				& $11.42\pm0.04$		& $5.2\pm1.0$\\
	& \ion{Mg}{II}	&	$0.0\pm0.1$			& $13.46\pm0.05$		& $6.9\pm0.3$\\
	& \ion{S}{II}	&	$0.0$				& $13.76^{+0.31}_{-0.27}$& $8.2^{+2.5}_{-1.6}$\\
	& \ion{S}{III}	&	$0.0$				& $14.22^{+0.12}_{-0.08}$& $11.0^{+5.4}_{-4.2}$\\
	& \ion{Fe}{II}	&	$0.0$				& $13.91\pm0.05$		& $4.8\pm0.2$ \\  \hline

5	&  \ion{H}{I}	&	$+23.9$				& $16.53^{+0.06}_{-0.10}$	& $18$  \\  
	& \ion{C}{II}	&	$+23.9$				& $>14.09$				& $<10.9$\\	
	& \ion{C}{III}	&	$+23.9$				& $>13.75$			  	& $<19.5$\\
	& \ion{N}{II}	&	$+23.9$				& $13.95^{+0.38}_{-0.36}$	& $5.3^{+2.1}_{-1.0}$\\
	& \ion{O}{II}	&	$+23.9$				& $<13.91$			  	& $10$\\
	& \ion{Mg}{I}	&	$+23.9$				& $<10.7$					& 10\\
	& \ion{Mg}{II}	&	$+23.9\pm0.1$			& $13.13\pm0.10$			& $3.7\pm0.3$\\
	& \ion{S}{II}	&	$+23.9$				& $13.70^{+0.38}_{-0.32}$	& $7.0^{+3.0}_{-1.8}$\\
	& \ion{S}{III}	&	$+23.9$				& $<13.3$					& $10$\\
	& \ion{Fe}{II}	&	$+23.9$				& $12.57\pm0.04$			& $3.1\pm0.9$\\  \hline

6	& \ion{H}{I}	&	$+42.3$				& $15.90\pm0.08$		& $15.7^{+1.5}_{-1.2}$  \\   
	& \ion{C}{II}	&	$+42.3$				& $14.16^{+0.17}_{-0.10}$&	$8.3^{+2.1}_{-1.6}$\\	
	& \ion{C}{III}	&	$+42.3$				& $>13.33$		& $<19.2$ \\
	& \ion{N}{II}	&	$+42.3$				& $13.22^{+0.20}_{-0.47}$		& $7.6^{+1.9}_{-1.1}$\\
	& \ion{O}{II}	&	$+42.3$				& $<14.41$			& 10\\
	& \ion{Mg}{I}	&	$+42.3$				& $<10.7$				& 10\\
	& \ion{Mg}{II}	&	$+42.3\pm0.2$			& $12.90\pm0.01$		& $6.6\pm0.3$\\
	& \ion{S}{II}	&	$+42.3$				& $<13.1$				& 10\\
	& \ion{S}{III}	&	$+42.3$				& $<13.3$				& $10$\\
	& \ion{Fe}{II}	&	$+42.3$				& $12.77\pm0.03$		& $4.8\pm0.6$\\  \hline
high-1	& \ion{O}{VI}	&	$-84.8\pm28.9$				& $14.54\pm0.04$			& $89.8\pm10.1$\\ 
high-2	& \ion{O}{VI}	&	$-26.9\pm34.4$				& $13.92\pm0.15$			& $32.5\pm15.3$\\  
high-3	& \ion{O}{VI}	&	$+25.7\pm12.3$			& $14.29\pm0.06$			& $28.2\pm3.6$\\  \hline

\end{tabular}}
\end{scriptsize}
\end{center}
\end{table}

\begin{table*}
\begin{center}
\caption{Summary of the ionization modeling results for the CUBS LLS sample.}
\hspace{-2.5em}
\label{tab:Imaging}
\resizebox{6.5in}{!}{
\begin{tabular}{@{\extracolsep{3pt}}cccccc@{}}\hline
Component	&$\mathrm{log\,}n_\mathrm{H}/\cmjjj$&$\mathrm{[\alpha/H]}$ 	& $\mathrm{[N/\alpha]}$ & $\mathrm{[Fe/\alpha]}$ & $\mathrm{[C/\alpha]}$ 	\\
	\\	\hline 

\multicolumn{6}{c}{J0248$-$4048 at $z_\mathrm{abs}=0.36400$, log\,$N$(\ion{H}{I})/\cmjj$=17.57\pm0.01$}	 \\ \hline
1&	$-2.89^{+0.14}_{-0.14}$	& $-1.36^{+0.12}_{-0.11}$	& $<1.15$				& $<1.50$				& $0.46^{+0.12}_{-0.13}$ \\
2&	$-1.83^{+0.13}_{-0.14}$	& $-0.87^{+0.05}_{-0.05}$	& $<-0.60$			& $0.26^{+0.15}_{-0.15}$	& $0.02^{+0.18}_{-0.19}$	\\	
3&	$-1.65^{+0.05}_{-0.05}$	& $-0.74^{+0.04}_{-0.14}$	& $-0.14^{+0.11}_{-0.11}$	& $0.38^{+0.05}_{-0.05}$	& $-0.09^{+0.07}_{-0.07}$	\\	
4&	$-3.21^{+0.39}_{-0.44}$	& $-0.89^{+0.43}_{-0.30}$	& $<-0.50$			& $<1.27$				& $<-0.34$\\	
5&	$-3.30^{+0.61}_{-0.67}$	& $<-0.09$			& $<1.10$				& $<2.54$				& $>-0.87$	\\	
6&	$-3.31^{+0.42}_{-0.51}$	& $-0.61^{+0.57}_{-0.46}$	& $<0.13$				& $<1.85$				& $>-1.09$	\\	\hline
\multicolumn{6}{c}{J0357$-$4812 at $z_\mathrm{abs}=0.43526$, log\,$N$(\ion{H}{I})/\cmjj$=17.18\pm0.01$}	 \\ \hline
1						& $-2.09^{+0.06}_{-0.06}$	& $-0.94^{+0.07}_{-0.07}$	& $0.08^{+0.14}_{-0.14}$	& $<0.42$				& $-0.14^{+0.08}_{-0.08}$ \\
2						& $-2.51^{+0.12}_{-0.11}$	& $-1.87^{+0.05}_{-0.05}$	& $-0.40^{+0.25}_{-0.25}$	& $<1.14$				& $0.09^{+0.21}_{-0.20}$	\\	
3\,(high density)			& $-2.41^{+0.11}_{-0.11}$	& $-0.64^{+0.08}_{-0.08}$	& $-0.01^{+0.18}_{-0.18}$	& $0.46^{+0.25}_{-0.25}$	& $0.66^{+0.25}_{-0.25}$	\\	
3\,(low density)				& $-3.92^{+0.16}_{-0.18}$	& $-0.57^{+0.42}_{-0.37}$	& $<0.79$				& $<2.73$				& $<2.67$	\\	
4						& $<-2.12$			& $<-0.25$			& $<1.74$				& $<2.71$				& $>0.10$\\	
5						& $<-2.59$			& $<-0.50$			& $<1.56$				& $<2.73$				& $>0.59$	\\	
6						& $-4.30^{+0.11}_{-0.11}$	& $-0.89^{+0.19}_{-0.17}$	& $<0.15$				& $<2.71$				& $0.35^{+0.28}_{-0.27}$	\\ \hline	
\multicolumn{6}{c}{J2135$-$5316 at $z_\mathrm{abs}=0.62255$, log\,$N$(\ion{H}{I})/\cmjj$=18.01\pm0.04$}	 \\ \hline
2&	$-2.78^{+0.28}_{-0.37}$	& $-0.19^{+0.61}_{-0.32}$	& $>-0.55$			& $<2.67$				& $>-0.63$	\\	
3&	$-0.69^{+0.27}_{-0.23}$	& $-0.93^{+0.05}_{-0.06}$	& $<0.62$				& $-0.50^{+0.07}_{-0.07}$	& $-0.10^{+0.25}_{-0.23}$	\\	
4&	$-1.98^{+0.26}_{-0.29}$	& $-0.16^{+0.09}_{-0.09}$	& $<0.11$				& $0.31^{+0.31}_{-0.25}$	& $0.34^{+0.22}_{-0.21}$\\	
5&	$<-3.56$				& $<-0.09$			& $<0.29$				& $<2.75$				& $>-0.28$	\\ \hline	
\multicolumn{6}{c}{J2308$-$5258 at $z_\mathrm{abs}=0.54273$, log\,$N$(\ion{H}{I})/\cmjj$=17.59\pm0.02$}	 \\  \hline
1&	$-2.18^{+0.15}_{-0.13}$	& $0.01^{+0.03}_{-0.03}$	& $0.55^{+0.17}_{-0.17}$	& $0.46^{+0.15}_{-0.15}$	& $0.24^{+0.34}_{-0.34}$ \\
2&	$<-3.45$				& $<0.83$				& $<1.66$				& $<2.69$				& $0.40^{+0.86}_{-0.62}$	\\	
3&	$-1.73^{+0.28}_{-0.26}$	& $-0.42^{+0.14}_{-0.09}$	& $0.09^{+0.26}_{-0.23}$	& $0.08^{+0.26}_{-0.25}$	& $>0.18$	\\	
4&	$-2.08^{+0.15}_{-0.17}$	& $-0.30^{+0.23}_{-0.28}$	& $>-0.45$			& $0.35^{+0.55}_{-0.39}$	& $>0.37$\\	
5&	$-1.30^{+0.18}_{-0.24}$	& $-0.03^{+0.06}_{-0.09}$	& $0.42^{+0.44}_{-0.40}$	& $-0.35^{+0.13}_{-0.13}$	& $>0.39$	\\	
6&	$-2.38^{+0.11}_{-0.14}$	& $0.15^{+0.09}_{-0.11}$	& $-0.23^{+0.33}_{-0.33}$	& $0.55^{+0.22}_{-0.15}$	& $0.25^{+0.21}_{-0.20}$	\\


\hline
\end{tabular}}
\end{center}
\end{table*}

\label{lastpage}

\end{document}